%% file: journal-R1-no-blue.tex
\documentclass[twocolumn]{IEEEtran}
\usepackage{amsmath,amssymb,amsthm,epsfig,color,subfigure,empheq,graphicx,graphics,balance}
\usepackage{enumerate,url,algorithm,algorithmic,wasysym,epstopdf,enumitem,array}
\usepackage{xcolor}
\usepackage{amsfonts}
\usepackage{adjustbox}
\usepackage{multirow}

\usepackage{accents}

\input{stylesV2}

\begin{document}

\title{Solving Optimal Power Flow on a Data-Budget:\\
Feature Selection on Smart Meter Data}

\author{Vassilis Kekatos,~\IEEEmembership{Senior Member,~IEEE}, 
        Ridley Annin,~\IEEEmembership{Student Member,~IEEE}, \\
        Manish K. Singh,~\IEEEmembership{Senior Member,~IEEE}, and 
	    Junjie Qin,~\IEEEmembership{Member,~IEEE}	
\vspace*{-2em}
}	
	
\markboth{IEEE TRANSACTIONS ON POWER SYSTEMS (submitted February 10, 2025; revised July 12, 2025)}{IEEE TRANSACTIONS ON POWER SYSTEMS (submitted February 10, 2025); revised July 12, 2025}

\maketitle

\begin{abstract}
How much data is needed to optimally schedule distributed energy resources (DERs)? Does the distribution system operator (DSO) have to know load demands at each bus of the feeder to solve an optimal power flow (OPF)? This work exploits redundancies in OPF's structure and data to minimize the communication of such a data deluge, and explores the trade-off between data compression and the grid's performance. We propose an OPF data distillation framework involving two steps: The DSO first collects OPF data from only a subset of nodes. It subsequently reconstructs the complete OPF data from the partial ones, and feeds them into the OPF solver. Selecting and reconstructing OPF data may be performed to maximize the fidelity of the reconstructed data or the associated OPF solutions. Under the first objective, OPF data distillation is posed as a sparsity-regularized convex problem. Under the second objective, it is posed as a sparsity-regularized bilevel program. Both problems are solved using proximal gradient algorithms. {The second objective is superior in approximating OPF solutions at the expense of increased complexity.} Numerical tests show that it enhances the fidelity and feasibility of the reconstructed OPF solutions, which can be approximated reasonably well even from partial data. 
\end{abstract}
	


\section{Introduction}
\allowdisplaybreaks
With the rampant integration of DERs, distribution system operators (DSOs) need to dispatch resources optimally at increasingly finer spatiotemporal scales. Solving the OPF requires collecting a large amount of data in near real-time. Currently, smart meters report packetized load demand data hourly or daily. Nonetheless, to control DERs effectively, the DSO may have to communicate with all customers and collect their active and reactive power demands every few minutes. This technical specification may be challenging to meet due to concerns related to communication, data privacy, and cyberattacks. This work leverages redundancies in OPF features and the structure of the OPF to approximately solve the OPF using only a carefully selected subset of load demands. {The primary objective is to reduce the real-time communication overhead between customers and the DSO. This is critical because although smart meters collect data at fine temporal scales, such data often arrives at the DSO's control room in batches after a few hours due to networking issues.}

In transmission systems, the operator estimates voltages from grid measurements, infers loads by plugging voltages into the power flow equations, and subsequently feeds the inferred load values into real-time market operations. This workflow may not be replicated in real-time distribution system operations due to a lack of observability at the grid edge. Under this setting, can DSOs control DERs based on partial grid data? The OPF can be viewed as a parametric optimization problem: Given load demands at all buses captured by vector $\btheta$, the OPF returns the optimal DER dispatch denoted by vector $\bx(\btheta)$. A key challenge in real-time DSO operations is that $\btheta$ may be uncertain or incomplete. The OPF mapping $\btheta\rightarrow \bx$ has been extensively studied in the literature; see~\cite{L2O2021,JSKGL22} and references therein. Uncertainty quantification studies how a probabilistic characterization postulated on $\btheta$ propagates through the OPF~\cite{UQ-OPF}. Sensitivity analysis computes the Jacobian matrix of $\bx$ with respect to $\btheta$; see~\cite[Ch.~6]{Conejo06}. Stochastic or robust programming aims at securing the system's performance when $\btheta$ is not precisely known~\cite{UncertainTutorial}. Reference~\cite{Mieth24} studies the value of data in grid operations and how obfuscation to preserve privacy can affect OPF solutions. {Schemes for securing privacy in smart meter data are devised in~\cite{VladDvorkin,Giaconi21,Erkin13}. The concept of differential privacy is adopted in \cite{9573387} to solve the OPF while ensuring the privacy of electricity customers.} 

The focus here is on compressing $\btheta$ to approximately solve the OPF. Kron/Ward reduction techniques from the 1950's and more recent ones aim at compressing the OPF problem itself via a reduced network model~\cite{9992730}, \cite{7872506}, \cite{6506059}. However, mapping results back to the actual system is non-trivial. {Compression has also been leveraged in the context of \emph{learning-to-optimize}, wherein a deep neural network (DNN) is trained to predict OPF solutions. References \cite{Mak21} and \cite{park2023compact} suggest compressing the input and output space of the OPF mapping, respectively, so that the OPF mapping can be learned by a DNN having fewer trainable weights. Reference~\cite{OPFandLearnTSG21} introduces an encoder-decoder DNN structure to design a semi-centralized DER control policy. Although jointly trained, the encoder is implemented at the DSO and the decoder at DERs. Because the encoder output is of limited dimension, DERs can be controlled by a data-frugal control signal issued by the DSO.}

{Instead of compressing $\btheta$ or $\bx$, this work aims at selecting a subvector of $\btheta$ to feed into the standard OPF module.} References~\cite{GKJ2020} and \cite{kotary24} suggest a joint learning and optimization approach, according to which OPF data are unknown and can only be predicted from primitive features $\bphi$, such as weather forecasts, historical trends, or partial OPF data. Rather than first inferring $\btheta$ from $\bphi$ and then mapping $\btheta$ to $\bx$ via the OPF or an ML model, references \cite{GKJ2020} and \cite{kotary24} advocate learning the mapping from $\bphi$ to $\bx$. In our setting, the DSO has no access to $\bphi$ and must decide which OPF data to sample in real time. Keeping the OPF module in the workflow ensures that data distillation does not affect the OPF solver, allowing the same OPF solver to be reused regardless of whether the DSO collects more or less data. 

The task of selecting relevant OPF data is reminiscent of \emph{sparse learning} or \emph{feature selection} in machine learning~\cite{yuan2007model}, wherein given \emph{feature vector} $\btheta$, one tries to infer a target variable $\bx(\btheta)$ using only a subset of features. Although feature selection and reconstruction have been extensively studied in the ML context, we revisit them with a fresh look under the OPF lens. Clearly, selecting OPF data is closely related to the optimal placement of meters or actuators. {Reference~\cite{BKVGS17} considers the task of inferring power injections at all buses of a distribution grid by polling smart meter data from only a subset of buses. The problem of installing actuators to improve controllability in transmission systems has been studied in~\cite{6740090}. Reference~\cite{TKC19} activates a subset of inverter-based resources to learn the distribution grid topology, and~\cite{buason2023datadriven} monitors a subset of buses in real time to ensure voltages remain within allowable limits. Perhaps most closely related to this work, reference \cite{2024rangarajanforecaste} considers selecting a subset of smart meters to poll in real time to improve distribution grid state estimation. Rather than improving the grid's observability or controllability, the goal here is to identify which data streams are most critical for approximating OPF solutions.}

The technical contributions of this work are on three fronts: 
\begin{itemize}
    \item[\emph{i)}] We formulate the task of OPF data distillation to select the most influential data to optimally schedule DERs on a data budget (Section~\ref{sec:problem});
    \item[\emph{ii)}] If the goal is to improve the fidelity of reconstructed OPF data, we adopt a greedy algorithm based on the discrete empirical interpolation method (DEIM) and a proximal gradient algorithm for solving a group-lasso-type convex optimization problem (Section~\ref{sec:data}); and
    \item[\emph{iii)}] If the goal is to improve the fidelity of reconstucted OPF solutions, we put forth a bilevel optimization program and develop a proximal gradient algorithm for solving it (Section~\ref{sec:opfdata}).
\end{itemize}
The developed methods are evaluated on a single-phase simplification of the IEEE 37-bus benchmark feeder {and a 1,136-bus feeder.} We solved prototypical examples of the OPF that control reactive power injections from DERs to minimize losses while maintaining bus voltages within allowable limits. Grid loading conditions were simulated using a combination of synthetic and real-world load, solar, and electric vehicle data. The tests corroborate that fidelity improves when selecting more OPF features, OPF-aware data distillation yields better results in terms of approximating optimal DER schedules and complying with constraints, and OPF solutions can be approximated reasonably well given only a relatively small subset of OPF data. Conclusions and open research directions are presented in Section~\ref{sec:conclusions}. Partial preliminary results have been reported in the conference precursor of this work~\cite{IREP2025}.

{\emph{Notation:} Column vectors (matrices) are denoted by lower-(upper-) case letters. Symbol $(\cdot)^\top$ stands for transposition; $\bI_N$ is the $N \times N$ identity matrix; $\|\bx\|_2$ is the $\ell_2$-norm of vector $\bx$; and $\|\bX\|_F$ is the Frobenius matrix norm.}

\section{Problem Statement}\label{sec:problem}
A distribution system operator (DSO) wants to decide the optimal setpoints for reactive power injections by inverter-interfaced DERs. To this end, the operator routinely collects data across a feeder, which is subsequently used as parameters in an OPF. Such data may be of a sizable volume, as they include active injections by DERs, as well as active and reactive demands from inflexible loads. Such data may be aggregated at the level of primary network buses. Under limited communication and/or sensing capabilities, the operator may not be able to possess all such data promptly to solve the OPF in real-time. In other words, if the OPF must be solved every few minutes, the operator may be unable to communicate and read data from every single bus of the primary network. A more realistic setup is that the operator can afford to install sensors only at $K$ out of the $N$ buses of a feeder. Hence, in real-time, the operator would communicate only with these $K$ sensors and rely solely on their data feeds to solve the OPF. Under this setup, two questions arise: \emph{i) How to sample data optimally?} and \emph{ii) How to solve the OPF using partial data?}

To address these questions, we deal with a prototypical OPF version. Consider the task of optimally deciding the reactive power setpoints by DERs. Given re/active loads and active solar generation on each bus, the DSO intends to decide optimally on DER reactive power setpoints. Such setpoints could be determined by solving an OPF that minimizes ohmic power losses on distribution lines while ensuring that bus voltages remain within allowable limits. The OPF considers a linearized grid model of {a single-phase radial test feeder.} Although the technical exposition considers this stylized version of the OPF, our proposed methodology applies readily to other settings. 

Consider a single-phase feeder with $N+1$ buses, indexed by $n=0,\ldots, N$. The substation bus is indexed by $n=0$. Let $v_n$ be the voltage magnitude, and $p_n+jq_n$ the complex power injection at bus $n$. Collect all but the substation injections and voltages in the $N$-length vectors $(\bp,\bq,\bv)$. Vectors $(\bp,\bq)$ can be decomposed into inverter injections $(\bp^g,\bq^g)$ and load withdrawals $(\bp^\ell,\bq^\ell)$ as
\begin{subequations}\label{eq:pq}
\begin{align}
\bp&=\bp^g-\bp^\ell\label{eq:pq:p}\\
\bq&=\bq^g-\bq^\ell.\label{eq:pq:q}
\end{align}
\end{subequations} 
Without loss of generality, each bus is assumed to host at most one (aggregate) load and at most one DER.

We adopt an approximate grid model rather than the exact AC power flow equations. Modeling inaccuracies can be justified due to the \emph{planning} rather than operational nature of our OPF data distillation task: The goal at this point is \emph{not} to schedule DERs, but to find which data are most valuable for scheduling DERs. The approximate grid model is obtained upon linearizing the power flow equations at the flat voltage profile of unit magnitudes and zero angles~\cite{TJKT20}. Then, nodal voltage magnitudes depend linearly on power injections as 
\begin{equation}\label{eq:voltage}
\bv(\bp,\bq) \simeq \bR\bp + \bX\bq + v_0\bone
\end{equation}
where matrices $(\bR,\bX)$ depend on the feeder topology and line impedances; $v_0$ is the substation voltage; and $\mathbf{1}_N$ is the vector of all-one entries. 

To formulate the OPF, we would also need to express ohmic losses on lines as functions of $(\bp,\bq)$. If we use the flat voltage profile again as the linearization point and pursue a first-order Taylor's series expansion, ohmic losses take the constant value of zero. If we use a second-order Taylor's series expansion, ohmic losses on all lines can be expressed as a convex quadratic function of power injections as~\cite{TJKT20}
\begin{equation}\label{eq:losses}
L(\bp,\bq)\simeq 2\bp^\top\bR\bp + 2\bq^\top\bR\bq.
\end{equation}

Based on \eqref{eq:voltage} and \eqref{eq:losses}, the optimal DER setpoints can be found as the solution to a convex quadratic program as
\begin{subequations}\label{eq:opf0}
	\begin{align}
	\min_{\bq^g}~&~\bq^\top\bR\bq\label{eq:opf0:cost}\\
	\mathrm{subject~to~(s.to)}&~-\bbv\leq \bR\bp+\bX\bq\leq \bbv\label{eq:opf0:c1}\\
	           &~-\bbq^g\leq \bq^g\leq \bbq^g~~\text{and}~~\eqref{eq:pq}\label{eq:opf0:c2}
	\end{align} 
\end{subequations}
where $\bbv=0.03\bone$. Problem~\eqref{eq:opf0} minimizes losses with respect to DER reactive setpoints $\bq^g$. Constraint~\eqref{eq:opf0:c1} maintains voltage deviations within the allowable range of $\pm3\%$~per unit (pu). To derive~\eqref{eq:opf0:c1}, it has been tacitly assumed that the substation voltage is $v_0=1$~pu, so that $\bv-v_0\bone\simeq\bR\bp+\bX\bq$ from \eqref{eq:voltage}. Alternatively, the substation voltage $v_0$ could be included as a decision variable, in which case, constraint \eqref{eq:opf0:c1} should be replaced by $0.97\cdot\bone\leq \bR\bp+\bX\bq+v_0\bone\leq 1.03\cdot\bone$. Moreover, voltage regulators (either remotely controllable or operating autonomously driven by local settings) can also be included in the OPF as in~\cite{STKSL22}. We skip both extensions to keep the presentation succinct. Constraint~\eqref{eq:opf0:c2} limits inverter setpoints within reactive power ratings collected in vector $\bbq_g$, assumed known to the DSO and fixed.

Under extreme loading conditions, limited VAR capacity, and/or imprecise OPF data ($\bp^g,\bp^\ell,\bq^\ell$), problem \eqref{eq:opf0} may become infeasible. In other words, it may be impossible for DERs to maintain voltages within the specified limits. To deal with this possibility and comply with industry practice, voltage constraints can be converted to \emph{soft constraints}: A slack variable $s\geq 0$ is added on voltage constraints and is penalized in the objective as
\begin{subequations}\label{eq:opf}
	\begin{align}
	\min_{\bq^g,s}~&~\bq^\top\bR\bq+ \ \nu s^2 +\rho s\label{eq:opf:cost}\\
	\mathrm{s.to}~&~-s\bone-\bbv\leq \bR\bp+\bX\bq\leq \bbv+s\bone\label{eq:opf:c1}\\
	           &~-\bbq^g\leq \bq^g\leq \bbq^g.\label{eq:opf:c2}
	\end{align} 
\end{subequations}
If the positive constants $(\nu, \rho)$ are selected sufficiently large, problem~\eqref{eq:opf} exhibits the following neat property~\cite{TJKT20}. For feasible instances of \eqref{eq:opf0}, problem~\eqref{eq:opf} yields optimal $s^*=0$ and the $\bq$-minimizers of the two problems coincide. On the other hand, for infeasible instances of \eqref{eq:opf0}, problem \eqref{eq:opf} returns $s^*>0$ and its $\bq$-minimizer satisfies voltage limits stretched out by $s^*$. In this way, problem~\eqref{eq:opf} deals with feasible and infeasible instances of the OPF systematically. Reference~\cite{TJKT20} provides analytical bounds and practical heuristics on how to select $\rho$ and $\nu$ appropriately.

\begin{figure*}[t]
    \centering
    \includegraphics[scale=0.4]{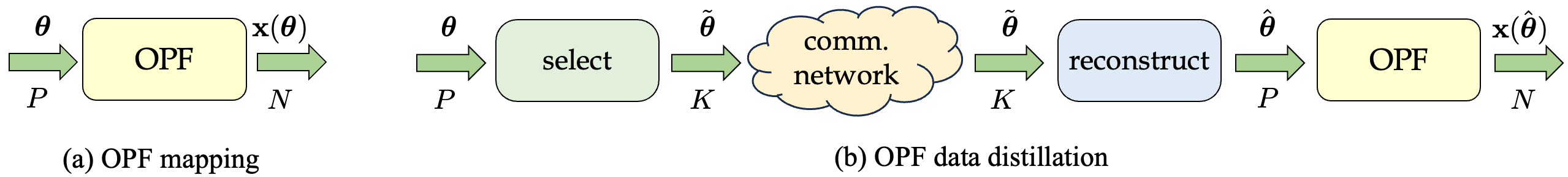}
    \caption{\emph{(a) OPF mapping:} The OPF as a parametric optimization problem. The OPF maps input data $\btheta\in\mathbb{R}^P$ to minimizers $\bx\in\mathbb{R}^N$. \emph{(b) OPF data distillation:} The operator first selects $K$ out of $P$ data to form $\tbtheta\in\mathbb{R}^K$. The entries of $\tbtheta$ are communicated from each selected bus to the operator over a communication network. The operator uses $\tbtheta$ to reconstruct an OPF parameter vector $\hbtheta$ of length $P$, equal to the original parameter vector $\btheta$. The operator feeds $\hbtheta$ into the OPF solver to find the DER dispatch $\bx(\hbtheta)$. The OPF solver remains the same. How to optimally design the \emph{selection function} (green) and the \emph{reconstruction function} (blue)? These two functions are designed and implemented offline and then used in real-time. Should they be designed to minimize $\|\btheta-\hbtheta\|$ or $\|\bx(\btheta)-\bx(\hbtheta)\|$? See Sections~\ref{sec:data} and \ref{sec:opfdata}, respectively.}
    \label{fig:bd}
\end{figure*}

Substituting \eqref{eq:pq} into \eqref{eq:opf} yields an optimization problem over $\bq^g$ and $s$. For simplicity, the minimizer of \eqref{eq:opf} will be henceforth denoted by $\bx=[\bq^g;~s]$ using a MATLAB-like notation. The OPF depends on $(\bR,\bX)$ and \emph{OPF data vector} 
\begin{equation}\label{eq:theta}
\btheta=\left[\bp^g-\bp^\ell;~~\bq^\ell\right].
\end{equation}
The DSO would like to solve \eqref{eq:opf} for different values of $\btheta$ and/or $(\bR,\bX)$. To keep the exposition uncluttered, we assume the feeder topology to remain fixed, and vary only the grid loading conditions $\btheta$. Either way, it is reasonable to assume $(\bR,\bX)$ to be known to the DSO, whereas $\btheta$ is the communication-intensive component of \eqref{eq:opf}. Even though $\btheta$ in \eqref{eq:theta} has been defined to be of dimension $2N$, some of the buses may be zero-injection buses, or their entries may be already known to the operator through other means. We will henceforth denote the actual length of $\btheta$ by $P$, where $P\leq 2N$. We slightly abuse notation and index the entries of $\btheta$ by subscript $p$ as $\theta_p$. We often denote the OPF minimizer as $\bx(\btheta)$ to emphasize its dependence on $\btheta$. To put it differently, the OPF is a \emph{parametric optimization problem} and can be interpreted as a mapping from $\btheta$ to $\bx$; see Fig.~\ref{fig:bd}(a). 

To solve~\eqref{eq:opf}, the DSO needs to know $\btheta$ in nearly real-time. This requirement may be impractical due to communication or privacy concerns. As a remedy, the operator may install sensors that read injections on a subset $K$ of $P$ data sources. If a bus is instrumented with a sensor, it is reasonable to assume the operator has real-time access to the corresponding entries of $(\bp_g-\bp_\ell)$ and $\bq_\ell$. The pertinent questions are: 
\begin{itemize}
    \item[\emph{q1)}] \emph{How to select $K$ out of $P$ OPF data sources?} 
    \item[\emph{q2)}] \emph{Once $K$ data sources have been selected, how to reconstruct a $P$-dimensional vector to feed into problem \eqref{eq:opf}?}
\end{itemize}

This task of \emph{OPF data distillation} is visualized in Fig.~\ref{fig:bd}(b). The operator selects $K$ out of $P$ entries of $\btheta$ to create vector $\tbtheta$. Given the selected OPF data $\tbtheta$, the DSO reconstructs a data vector $\hbtheta$ to feed into the OPF. The reconstructed vector $\hbtheta$ should be of the original dimension $P$. This ensures backward compatibility with the OPF solver and keeps the solver independent of the data distillation scheme, which may change with time.

An OPF data distillation scheme involves \emph{data selection} and \emph{data reconstruction}. The selection step can be expressed as $\tbtheta=\bS^\top\btheta$, where $\bS$ is a selection matrix with a single one-entry per column and zeros elsewhere. Vector $\tbtheta$ is a subvector of $\btheta$. If we limit ourselves to linear schemes, the reconstruction step can be expressed with the help of a $P\times K$ matrix $\bC$ as $\hbtheta=\bC\tbtheta$. Combining the two steps into $\bW=\bC\bS^\top$ yields:
\begin{equation}\label{eq:distil}
\hbtheta=\bC\tbtheta=\bC\bS^\top \btheta=\bW\btheta.
\end{equation}

OPF data distillation amounts to finding $(\bC,\bS)$. These two matrices can be designed offline using a dataset of $T$ grid loading scenarios of OPF data, arranged as columns of the $P\times T$ matrix
\begin{equation}\label{eq:Theta}
\bTheta=[\btheta_1~\btheta_2~\cdots~\btheta_T].
\end{equation}
Once designed, matrices $(\bC,\bS)$ can be held constant over a longer operational period depending on the application or until the statistics of OPF data change from the statistics of $\bTheta$.

The training OPF data have been centered and scaled so each row of $\bTheta$ is zero-mean and unit-variance. Such preprocessing is necessary to render all data sources tentatively equally important. Otherwise, an almost time-invariant load demand of 1~pu may seem more important than a highly time-variable load demand of 0.5~pu. All subsequent approaches operate on the preprocessed normalized OPF data $\bTheta$. Once normalized OPF data have been distilled, the data fed into the OPF are returned to their original (non-normalized) form. 


Matrices $(\bC,\bS)$ can be designed to optimize a meaningful metric. A first idea is to design $(\bC,\bS)$ so that $\hbtheta$ approximates $\btheta$ well. Such distillation schemes focus on the fidelity of \emph{OPF data} and will be henceforth referred to as \emph{Type-1 schemes}. A second idea is to design $(\bC,\bS)$ so that $\bx(\hbtheta)$ approximates $\bx(\btheta)$ well. Such schemes focus on the fidelity of \emph{OPF decisions} and will be referred to as \emph{Type-2 schemes}. Type-1 schemes are developed in Section~\ref{sec:data} and Type-2 schemes in Section~\ref{sec:opfdata}. Before presenting the proposed data distillation schemes, three remarks are in order.

\begin{remark}\label{re:comparison}
If the ultimate goal for the DSO is to find near-optimal OPF decisions using partial data, Type-2 schemes are preferable over Type-1 alternatives. Nonetheless, as it will become clear later, Type-2 schemes are computationally more demanding than Type-1 schemes. The performance advantage of Type-2 over Type-1 schemes is expected to diminish for larger data compression ratios $K/P$. Specifically, as the ratio $K/P$ tends to unity, the reconstructed OPF data $\hbtheta$ approaches the actual data $\btheta$, and thus, the approximate decisions $\bx(\hbtheta)$ are expected to lie close to $\bx(\btheta)$. In such cases, Type-1 schemes may perform reasonably well, forgoing the complexity of Type-2 schemes. On the other hand, for smaller $K/P$ ratios, the reconstructed data $\hbtheta$ may be away from $\btheta$, in which case the increased complexity of Type-2 schemes pays off in recovering better OPF decisions. 
\end{remark}

\begin{remark}\label{re:privacy+security}
The primary motivation for OPF data distillation is to reduce the real-time communication overhead on the uplink between electricity customers and the DSO's control room. Through the proposed schemes, the DSO communicates with only $K$ customers instead of $P$. Nonetheless, the proposed schemes could also be used to improve data privacy and cybersecurity issues as outlined next. Regarding data privacy, the DSO may ask customers if they consent to their real-time demand being used for scheduling purposes and select to sample only from those customers. Alternatively, the DSO may use data distillation schemes to identify data-critical customers and solicit their approval for data usage by providing discounts. Regarding cybersecurity, data distillation schemes may be beneficial in pinpointing the most critical data streams to protect by implementing higher cybersecurity protocols.
\end{remark}

\begin{remark}\label{re:general}
Although the OPF data distillation schemes proposed in the following sections refer to the specific OPF model in \eqref{eq:opf}, the suggested methodology is quite general. It can be readily applied to various OPF models, including AC OPF variants in single- or multi-phase grids. More specifically, Type 1 schemes are invariant to the OPF model. Even though Type 2 schemes depend on the OPF model, the only requirement is to be able to differentiate through the OPF mapping and compute $\nabla_{\btheta}\bx$. In previous works~\cite{L2O2021,JSKGL22,TJKT20}, we have shown that this Jacobian matrix can be easily computed for various OPF formulations, including linear or quadratic programs (LP/QP) relying on linearized grid models, and second-order cone or semidefinite programs (SOCP/SDP) relying on nonlinear AC grid models. Regardless of the model, computing $\nabla_{\btheta}\bx$ entails solving a system of linear equations, granted the optimal primal/dual OPF solutions have already been computed. 
\end{remark}
\color{black}

\section{Type-1 OPF Data Distillation}\label{sec:data}
This section presents Type-1 schemes so that distilled data $\hbtheta$ approximates well original data $\btheta$ in the training dataset. 

\subsection{OPF Data Compression using PCA}\label{subsec:pca}
Given dataset $\bTheta$, the first attempt could be to apply a dimensionality-reduction technique, such as principal component analysis (PCA). This seems intuitive as some entries of $\btheta_t$ may exhibit linear dependencies, e.g., due to solar irradiance. To this end, we perform an eigenvalue decomposition (EVD) on the \emph{data covariance matrix}
\begin{equation}\label{eq:C}
\bC_{\theta}=\frac{1}{T}\bTheta\bTheta^\top=\bU\bLambda\bU^\top
\end{equation}
and collect the eigenvectors related to the largest $K$ eigenvalues as columns of matrix $\bU_K$. A `compressed' version of $\btheta_t$ can be obtained as $\tbtheta=\bU_K^\top\btheta_t$, and a reconstructed one as
\begin{equation}\label{eq:pca0}
\hbtheta_t=\bU_K\tbtheta_t=\bU_K\bU_K^\top\btheta_t=\bW_\text{PCA}\btheta_t
\end{equation}
where $\bW_{\text{PCA}}=\bU_K\bU_K^\top$ is a projection matrix and is known to be the minimizer of
\begin{equation}\label{eq:pca}
\min_{\bW}\{f_1(\bW):\rank(\bW)=K\}
\end{equation}
where $f_1(\bW)$ is the data fitting cost over the training dataset
\begin{align}\label{eq:f1}
f_1(\bW)&=\frac{1}{2T}\sum_{t=1}^T\|\btheta_t-\hbtheta_t\|^2\nonumber\\
&=\frac{1}{2T}\sum_{t=1}^T\|\btheta_t-\bW\btheta_t\|^2\nonumber\\
&=\frac{1}{2T}\|\bTheta-\bW\bTheta\|_F^2.
\end{align}
Although $\bW_{\text{PCA}}\bTheta$ is the best rank-$K$ approximation of $\bTheta$ in the Frobenius norm, matrix $\bU_K^\top$ cannot be used for data selection because it is dense in general. Then, computing $\tbtheta_t=\bU_K^\top\btheta_t$ would require knowing the entire $\btheta_t$. To put it differently, matrix $\bW_{\text{PCA}}$ does not comply with the format $\bC\bS^\top$ of \eqref{eq:distil}. However, the reconstruction error $\bW_{\text{PCA}}$ attains
\begin{equation}\label{eq:pcaerror}
f_1(\bW_{\text{PCA}})=\frac{1}{2T}\|\bTheta-\bW_{\text{PCA}}\bTheta\|_F^2
\end{equation}
serves as a benchmark for Type-1 data distillation schemes as PCA minimizes $f_1(\bW)$ over all rank-$K$ matrices.

\subsection{OPF Data Selection using DEIM}\label{subsec:deim}
To achieve data selection, we propose using the \emph{discrete empirical interpolation method} (DEIM)~\cite{barrault2004empirical}. By design, DEIM complies with the format of~\eqref{eq:distil}. Given a selection matrix $\bS$, DEIM sets matrix $\bC$ in \eqref{eq:distil} as $\bC=\bU_K(\bS^\top\bU_K)^{-1}$, where $\bU_K$ carries again the top-$K$ eigenvectors of $\bC_\theta$ from \eqref{eq:pca0} and assuming $\bS^\top\bU_K$ is invertible. DEIM reconstructs data as
\begin{equation}\label{eq:deim}
\hbtheta_t=\bU_K(\bS^\top\bU_K)^{-1}\tbtheta_t=\bW_{\text{DEIM}}\btheta_t
\end{equation}
where $\bW_{\text{DEIM}}=\bU_K(\bS^\top\bU_K)^{-1}\bS^\top$ is a projection matrix as $\bW_{\text{DEIM}}^2=\bW_{\text{DEIM}}$. Although both $\bW_{\text{PCA}}$ and $\bW_{\text{DEIM}}$ are rank-$K$ projection matrices, the latter comes with the neat feature that $\hbtheta_t$ is computed using \emph{only} $K$ entries of $\btheta_t$. Moreover, for every vector in the dataset, it holds that $\bS^\top\hbtheta_t=\bS^\top\btheta_t$. In other words, the reconstructed $\hbtheta_t$ perfectly matches \emph{(interpolates)} $\btheta_t$ at the $K$ selected entries. On the other hand, matrix $\bW_{\text{DEIM}}$ is clearly suboptimal in terms of solving \eqref{eq:pca}. Therefore, its data fitting cost is lower bounded as $f_1(\bW_{\text{DEIM}})\geq f_1(\bW_{\text{PCA}})$. Interestingly enough, DEIM's data fitting cost is also upper bounded as~\cite{barrault2004empirical} 
\begin{equation}\label{eq:bound}
f_1(\bW_{\text{PCA}})\leq f_1(\bW_{\text{DEIM}}) \leq \|(\bS^\top\bU_K)^{-1}\|^2  f_1(\bW_{\text{PCA}})  
\end{equation}
where $\|\cdot\|$ is the spectral matrix norm. This bound holds for all selection matrices $\bS$ with invertible $\bS^\top\bU_K$. Since $\bU_K$ is found via EVD, matrix $\bS$ can be selected to minimize $\|(\bS^\top\bU_K)^{-1}\|$ and bring $f_1(\bW_{\text{DEIM}})$ closer to $f_1(\bW_{\text{PCA}})$. 

Nonetheless, minimizing $\|(\bS^\top\bU_K)^{-1}\|$ over $\bS$ is non-trivial. As a practical solution, DEIM has a greedy algorithm for selecting $\bS$ to minimize approximately $\|(\bS^\top\bU_K)^{-1}\|$; see~\cite{barrault2004empirical} for details. Bounds on the suboptimality of DEIM's greedy algorithm have been derived in~\cite{barrault2004empirical}. Empirical tests on different application domains demonstrate that DEIM performs significantly better than the analytically derived bounds. Heed that DEIM's greedy algorithm first requires computing $\bU_K$ by applying an EVD on $\bC_{\theta}$. Critically, the computational complexity of the greedy algorithm does not exceed the complexity of EVD. Therefore, DEIM has the same complexity order as computing the leading $K$ EVD components. 

\subsection{OPF Data Selection using Group Lasso}\label{subsec:glasso}
As an alternative to DEIM, we next formulate OPF data distillation as a type of a \emph{group lasso (GL) problem}~\cite{yuan2007model}. The reason for introducing yet another method is to bring more flexibility to Type-1 and Type-2 designs, as will be clear soon. Our GL-based, Type-1 data distillation entails solving the convex program:
\begin{equation}\label{eq:glasso}
\bW_{\text{GL}}=\arg\min_{\bW}f_1(\bW)+\lambda_1 g(\bW)
\end{equation}
where $g(\bW)=\sum_{p=1}^P\|\bw_p\|$, vector $\bw_p$ is the $p$-th column of $\bW$, and $\lambda_1>0$ is a tunable parameter.

The cost in~\eqref{eq:glasso} consists of two terms. The first term is the data fitting cost of PCA. The second term $g(\bW)$ is a regularization term that promotes column-sparse solutions~\cite{yuan2007model}. In other words, the minimizer $\bW_{\text{GL}}$ will have many zero columns with increasing $\lambda_1$. Expanding~\eqref{eq:distil}, we get that the reconstructed vector is 
\begin{equation}
\hbtheta = \sum_{p=1}^P\theta_p\bw_p.
\end{equation}
If column $\bw_p=\bzero$, OPF feature $\theta_p$ does not contribute to $\hbtheta$. 

GL offers more flexibility over DEIM. According to~\eqref{eq:theta}, active injection $p_n=p_n^g-p_n^\ell$ at bus $n$ corresponds to the $n$-th entry of $\btheta$, while its reactive injection $q_n=-q_n^\ell$ corresponds to the $(n+N)$-th entry of $\btheta$. Designing $\bW$ according to \eqref{eq:glasso} may give us $\bw_n\neq \bzero$ and $\bw_{n+N}=\bzero$. The same could happen with DEIM. Such design means the operator collects $p_n$, but not $q_n$. This may be impractical. If the operator is to install a sensor and communicate with bus $n$, they might as well either select both $(p_n,q_n)$ or none. This specification can be satisfied by GL (yet not DEIM) by simply placing columns $\bw_n$ and $\bw_{n+N}$ into the same group and replacing norms
\begin{equation}
\|\bw_n\|+\|\bw_{n+N}\|\quad \longrightarrow\quad\left\|\left[\begin{array}{l}\bw_n\\ \bw_{n+N}\end{array}\right]\right\|  
\end{equation}
in $g(\bW)$. GL offers additional flexibility as it can be modified to yield Type-2 designs, as discussed later in Section~\ref{sec:opfdata}.



Although~\eqref{eq:glasso} can be solved using interior point-based solvers as a second-order cone program (SOCP), such solvers do not scale favorably with the problem size. To improve computational complexity, we devise a first-order algorithm called \emph{accelerated proximal gradient} (APG) to solve~\eqref{eq:glasso}. Even though an APG algorithm has been developed for the general GL problem before~\cite{pgd4gl}, it is adopted to \eqref{eq:glasso} and reviewed here to ease the exposition of Type-2 schemes. 


APG is an extension of accelerated gradient descent when the cost of an optimization problem consists of a differentiable and a non-differentiable term~\cite{Li2015AcceleratedPG}. For our task in~\eqref{eq:glasso}, the differentiable term is $f_1(\bW)$, and the non-differentiable term is $g(\bW)$. Let APG iterations be indexed by $i$ and $\bW^i$ denote the estimate of minimizer $\bW$ at iteration $i$. Each iteration involves four steps~\cite{Li2015AcceleratedPG}:
\begin{subequations}\label{eq:pgd}
\begin{align}
\bbW^i&=\bW^i +\frac{\alpha_{i-1}-1}{\alpha_{i}}\left(\bW^i-\bW^{i-1}\right)\label{eq:pgd:a}\\
\bY^i&=\bbW^i-\mu_i \nabla_{\bW}f_1(\bbW^i)\label{eq:pgd:b}\\
\bW^{i+1}&=\arg\min_{\bW}~\lambda_1g(\bW)+\frac{1}{2\mu_i}\|\bW-\bY^i\|_F^2\label{eq:pgd:c}\\
\alpha_{i+1}&=\frac{1+\sqrt{4\alpha_i^2+1}}{2}\label{eq:pgd:d}
\end{align}
\end{subequations}
Step~\eqref{eq:pgd:b} coincides with a gradient descent step if we were to minimize $f_1(\bW)$ alone. It uses $\mu_i>0$ as a step size. However, the gradient descent step is not performed at $\bW^i$ but $\bbW^i$, computed as a linear combination of $\bW^i$ and $\bW^{i-1}$ in step~\eqref{eq:pgd:a}. The linear combination coefficients change across iterations according to~\eqref{eq:pgd:d}. After some mundane algebra, the gradient needed in \eqref{eq:pgd:b} can be found to be
\begin{equation}\label{eq:gradient}
\nabla_{\bW}f_1(\bW)=(\bW-\bI)\bC_{\theta}
\end{equation}
where $\bC_{\theta}$ is the OPF data covariance defined in \eqref{eq:C}.

\begin{algorithm}[t]
	\caption{APG for Type-1 OPF Data Distillation}\label{alg:PGD}
	\begin{algorithmic}[1]
		\renewcommand{\algorithmicrequire}{\textbf{Input:}}
		\renewcommand{\algorithmicensure}{\textbf{Output:}} 
		\REQUIRE OPF input data $\{\btheta_t\}_{t=1}^T$ stored in matrix $\bTheta$.
		\ENSURE OPF data distillation matrix $\bW$.
		\STATE Initialize $\bW^1=\bW^0$ from standard normal distribution.
        \STATE Initialize $\alpha_1=1$ and $\alpha_0=0$; set $\mu$ and $\lambda_1$.
		\FOR{$i = 1, 2, \ldots,$}
		\STATE Compute $\bY^i$ from \eqref{eq:pgd:a}--\eqref{eq:pgd:b} via \eqref{eq:gradient}.
		\STATE Compute the groups of $\bW^{i+1}$ from \eqref{eq:pgd:c} via \eqref{eq:prox}. 
        \STATE Compute $\alpha_{i+1}$ from \eqref{eq:pgd:d}.
        \ENDFOR
	\end{algorithmic}
\end{algorithm}

The output $\bY^i$ of the gradient descent step is then applied to the \emph{proximal operator} on $\bY^i$ to get $\bW^{i+1}$. The proximal operator is defined as the solution to the optimization in~\eqref{eq:pgd:c}. Problem~\eqref{eq:pgd:c} can be solved separately for each $\bw_p$. In fact, each block $\bw_p$ can be updated in closed form as~\cite{pgd4gl}:
\begin{equation}\label{eq:prox}
\bw_p^{i+1}=\mathrm{prox}\left(\by_p^i;\lambda_1\mu_i\right).
\end{equation}
Operator $\mathrm{prox}$ takes a vector $\bx$ and a scalar $\beta$, and returns:
\begin{equation}\label{eq:proxdef}
\mathrm{prox}\left(\bx;\beta\right)=\left\{
\begin{array}{ll}
\left(1-\frac{\beta}{\|\bx\|}\right)\bx&,~\|\bx\|\geq \beta\\
\bzero&,~\text{otherwise}.
\end{array}
\right.
\end{equation}
From \eqref{eq:prox}--\eqref{eq:proxdef}, if $\|\by_p^i\|\geq \lambda_1\mu_i$, vector $\bw_p^{i+1}$ is set to a scaled-down version of $\by_p^{i}$; otherwise, it is set to zero. This hints at why $g(\bW)$ effects group sparsity on $\bW_\text{GL}$. Step~\eqref{eq:pgd:b} applies regardless of whether the entries of $\bW$ are grouped in columns or other groups as long as groups do not share common entries. The APG algorithm is tabulated as Algorithm~\ref{alg:PGD}. If $\nabla_{\bW}f_1$ is $L$-smooth, the APG iterations with constant step size $\mu_i=\mu\leq 1/L$ converge to the minimum cost at a rate of $\mcO(1/i^2)$; see~\cite{Li2015AcceleratedPG}.

Some comments on $\lambda_1$ are due. For $\lambda_1=0$, the GL solution $\bW_{\text{GL}}$ coincides with $\bW_{\text{PCA}}$, which is dense in general. As $\lambda_1$ increases, more columns of $\bW_{\text{GL}}$ become zero monotonically. Lemma~\ref{le:lambda_max} provides an upper bound $\bar{\lambda}_1$ on $\lambda_1$ beyond which the minimizer of \eqref{eq:glasso} becomes zero; see the appendix for a proof. By solving \eqref{eq:glasso} for varying $\lambda_1\in[0,\bar{\lambda}_1]$, we can make $\bW_{\text{GL}}$ have exactly $K$ non-zero columns using bisection.

\begin{lemma}\label{le:lambda_max}
Let $\bar{\lambda}_1=\max_p \|\bc_p\|$, where $\bc_p$ is the $p$-th column of covariance matrix $\bC_{\theta}$. The zero matrix is a minimizer of \eqref{eq:glasso} if and only if $\lambda_1\geq\bar{\lambda}_1$.
\end{lemma}

The GL approach is known to successfully identify the sparsity support (non-zero columns) of $\bW$. It is nonetheless also known that the GL solution is a biased estimate of the column-sparse $\bW$ yielding the minimum fitting cost $f_1(\bW)$ alone. This is well understood for the general GL setting; our numerical tests indicate that the claim carries over to the OPF distillation task. As a remedy, we can employ a \emph{two-stage GL approach}: First, use GL only to select $K$ OPF features (columns of $\bW$). Based on the sparsity pattern of $\bW_\text{GL}$, we determine the selection matrix $\bS$ in \eqref{eq:distil} to the value of $\bS_\text{GL}$. Second, design the reconstruction matrix $\bC$ via a standard least-squares fit as:
\begin{align}
\bC_\text{GL2}&=\arg\min_{\bC}\|\bTheta-\bC\bS_\text{GL}^\top\bTheta\|_F^2\nonumber\\
&=\bTheta\bTheta^\top \bS_\text{GL} (\bS_\text{GL}^\top\bTheta\bTheta^\top\bS_\text{GL})^{-1}\nonumber\\
&=\bC_\theta \bS_\text{GL}(\bS_\text{GL}^\top\bC_\theta\bS_\text{GL})^{-1}.
\end{align}
The two-stage GL (GL2) returns the distillation matrix:
\begin{equation}
\bW_\text{GL2}=\bC_\text{GL2}\bS_\text{GL}^\top=\bC_\theta \bS_\text{GL}(\bS_\text{GL}^\top\bC_\theta\bS_\text{GL})^{-1}\bS_\text{GL}^\top.
\end{equation}

Type-1 methods are agnostic to the fact that distilled OPF data will eventually be used to produce OPF solutions and that we are primarily interested in OPF solutions rather than OPF data per se. To this end, we next explore Type-2 approaches. 

\section{Type-2 OPF Data Distillation}\label{sec:opfdata}
The ultimate goal of replacing $\btheta$ by $\hbtheta=\bW\btheta$ is \emph{not} to approximate $\btheta$ closely. The goal is to design $\hbtheta$ so that when fed into the OPF, it yields an OPF minimizer close to the original minimizer. To develop Type-2 designs, suppose the operator has access to a dataset $\mcD=\{(\btheta_t,\bx_t)\}_{t=1}^T$ of $T$ representative OPF scenarios paired with their OPF minimizers. 

We use the shorthand notation $\bx_t=\bx(\btheta_t)$ to denote the OPF minimizer of \eqref{eq:opf} given $\btheta_t$. We use a similar notation for the OPF minimizer given the reconstructed data as
\begin{equation}\label{eq:hbx}
\hbx_t=\bx(\hbtheta_t)=\bx(\bW\btheta_t)~~~\text{for}~t=1,\ldots,T.    
\end{equation}

We seek a column-sparse $\bW$ so that $\hbx_t$ is close to $\bx_t$ for all $t$. Paralleling~\eqref{eq:f1}, we define the \emph{OPF fitting error}:
\begin{equation}\label{eq:f2}
f_2(\bW)=\frac{1}{2T}\sum_{t=1}^T \|\bx_t-\hbx_t(\bW)\|^2.
\end{equation}
Minimizing $f_2(\bW)$ over a column-sparse $\bW$ is not amenable to a DEIM-type approach, but to a GL-type approach. We propose a Type-2 GL approach that we term \emph{bilevel group lasso} (BGL) that finds $\bW$ as the minimizer of:
\begin{subequations}\label{eq:bglasso}
\begin{align}
\bW_\text{BGL}=\arg\min_{\bW}&~F_2(\bW)=f_2(\bW)+\lambda_2 g(\bW)\label{eq:bglasso:cost}\\
\mathrm{s.to}&~\hbx_t~\text{minimizes}~\eqref{eq:opf}~\text{for}~\hbtheta_t=\bW\btheta_t~\forall t.\label{eq:bglasso:con}
\end{align}    
\end{subequations}
Like GL, the minimizer $\bW_\text{BGL}$ becomes more column-sparse for larger values of $\lambda_2>0$. The fitting term $f_2(\bW)$ aims at bringing $\hbx_t$ close to $\bx_t$. Recall that $\bx_t$'s have been already computed and given as part of dataset $\mcD$. On the other hand, each $\hbx_t$ is the minimizer of the OPF fed by $\hbtheta_t=\bW\btheta_t$, and so $\hbx_t$ depends on $\bW$. This is a \emph{bilevel optimization program}. The total cost $F_2(\bW)$ of the outer optimization over $\bW$ is determined by the minimizers of $T$ inner optimization problems over $\bx_t$'s. Each inner problem depends parametrically on $\bW$.

Problem~\eqref{eq:bglasso} is non-convex. Bilevel programs are often handled by replacing every inner problem with its optimality conditions~\cite{KKT2019}. Unfortunately, such an approach would lead to a large-scale mixed-integer nonlinear program (MINLP) that scales unfavorably with $N$ and $T$. Motivated by advancements in meta-learning, we solve \eqref{eq:bglasso} using an APG algorithm. Unfortunately, when it comes to non-convex problems, the standard APG algorithm is not guaranteed to converge to a \emph{critical point}, that is a point where the cost subdifferential contains zero (zero-gradient point if the cost is differentiable). To resolve this issue, we use a modification of APG proposed in~\cite[Alg.~2]{Li2015AcceleratedPG}. Although the modified APG algorithm is slightly more complex, it is guaranteed to converge to a critical point even for non-convex problems. The method is tabulated as Algorithm~\ref{alg:NAPG}. Its differences to Alg.~\ref{alg:PGD} are discussed next.

Compared to Algorithm~\ref{alg:PGD}, Algorithm~\ref{alg:NAPG} first updates $\bbW^i$ using three rather than two points as shown in Step~\ref{alg:NAPG:step5}. It then performs a gradient descent step based on $\nabla_{\bW}f_2$ evaluated at $\bbW^i$. The gradient $\nabla_{\bW}f_2$ can be computed as explained next and established in the appendix. 


\begin{algorithm}[t]
	\caption{APG for Type-2 OPF Data Distillation}\label{alg:NAPG}
	\begin{algorithmic}[1]
		\renewcommand{\algorithmicrequire}{\textbf{Input:}}
		\renewcommand{\algorithmicensure}{\textbf{Output:}} 
		\REQUIRE Dataset of OPF input/minimizers $\mcD=\{(\btheta_t,\bx_t)\}_{t=1}^T$.
		\ENSURE OPF data distillation matrix $\bW$.
        \STATE Initialize $\bZ^1=\bW^1=\bW^0$ and evaluate $c_1=F_2(\bW^1)$.
        \STATE Initialize $\alpha_1=1$, $\alpha_0=0$, $\eta\in [0,1)$, $\delta> 0$, and $q_1=1$.
		\STATE Select step sizes $\mu$ and $\bar{\mu}$, and parameter $\lambda_2$.
		\FOR{$i = 1, 2, \ldots,$}
		\STATE Update $\bbW^i=\frac{\alpha_i -1}{\alpha_i}\bW^i +\frac{\alpha_{i-1}}{\alpha_{i}}\bZ^i -\frac{\alpha_{i-1}-1}{\alpha_{i}}\bW^{i-1}$.\label{alg:NAPG:step5}
        \STATE Run the OPF for reconstructed data $\bbTheta^i = \bbW^i\bTheta$.\label{alg:NAPG:step6}
        \STATE Update $\bbY^i$ as $\bbY^i=\bbW^i-\bar{\mu}_i \nabla_{\bbW}f_2(\bbW^i)$.\label{alg:NAPG:step7}
        \STATE Update blocks of $\bZ^{i+1}$ as $\bz_p^{i+1}=\mathrm{prox}\left(\bby_p^i;\lambda_2\bar{\mu}_i\right)$.\label{alg:NAPG:step8}
        \IF{$F_2(\bZ^{i+1})\leq c_i -\delta\|\bZ^{i+1}-\bbW^i\|_F^2,$}\label{alg:NAPG:step9}
        \STATE $\bW^{i+1}=\bZ^{i+1}$.
        \ELSE
        \STATE Run the OPF for reconstructed data $\hbTheta^{i} = \bW^i\bTheta$.\label{alg:NAPG:step12}
        \STATE Update $\bY^i$ as $\bY^i=\bW^i-\mu_i \nabla_{\bW}f_2(\bW^i)$.\label{alg:NAPG:step13}
        \STATE Update blocks of $\bbZ^{i+1}$ as $\bbz_p^{i+1}=\mathrm{prox}\left(\by_p^i;\lambda_2\mu_i\right)$.\label{alg:NAPG:step14}
        \STATE $\bW^{i+1}=\begin{cases}
        \bZ^{i+1} &,~\text{if } F_2(\bZ^{i+1})\leq F_2(\bbZ^{i+1})\label{alg:NAPG:step15} \\
        \bbZ^{i+1} &,~\text{otherwise.} 
        \end{cases}$
        \ENDIF
        \STATE Update $\alpha_{i+1}=\left(1+\sqrt{4\alpha_i^2+1}\right)/2$.\label{alg:NAPG:step17}
        \STATE Update $q_{i+1}=\eta q_i +1$.\label{alg:NAPG:step18}
        \STATE Update $c_{i+1}=\frac{\eta q_i c_i+F_2(\bW^{i+1})}{q_{i+1}}$.\label{alg:NAPG:step19}
        \ENDFOR
	\end{algorithmic}
\end{algorithm}

\begin{lemma}\label{le:gradient}
Let $\nabla_{\hbtheta_t}\hbx_t$ be the Jacobian of the OPF minimizer $\hbx_t$ given reconstructed OPF data $\hbtheta_t=\bW\btheta_t$ for all $t$, then
\begin{equation}
\nabla_\bW f_2(\bW)=\frac{1}{T}\sum_{t=1}^T(\nabla_{\hbtheta_t}\hbx_t)^\top(\hbx_t-\bx_t)\btheta_t^\top.
\end{equation}
\end{lemma}

To take the gradient descent step, Algorithm~\ref{alg:NAPG} needs to solve $T$ OPFs using the reconstructed data $\bbtheta_t^i=\bbW^i\btheta_t$ computed from Step~\ref{alg:NAPG:step6}. Along with the minimizer, for each OPF, Algorithm~\ref{alg:NAPG} must compute the Jacobian matrix of the OPF minimizer with respect to the given OPF data. For Step~\ref{alg:NAPG:step6}, the minimizer is $\bbx_t=\bx(\bbtheta_t)$ and the OPF data is $\bbtheta_t$. Matrix $\nabla_{\bbtheta_t}\bbx_t$ is an $N\times P$ matrix carrying the sensitivities (partial derivatives) of $\bbx_t$ with respect to $\bbtheta_t$. Given the optimal primal/dual solutions of the OPF, matrix $\nabla_{\bbtheta_t}\bbx_t$ can be readily computed by solving of system of linear equations~\cite{SGKCB2020,L2O2021}.

Subsequently, Algorithm~\ref{alg:NAPG} applies the proximal operator on $\bbY^i$ to compute $\bZ^{i+1}$ in Step~\ref{alg:NAPG:step8}. Algorithm~\ref{alg:PGD} would  complete the iteration here and return $\bZ^{i+1}$ as $\bW^{i+1}$. Algorithm~\ref{alg:NAPG} does so only if $\bZ^{i+1}$ attains a sufficient decrease in total cost as detailed in Step~\ref{alg:NAPG:step9}. Note that evaluating $F_2(\bZ^{i+1})$ entails solving another batch of $T$ OPFs given data $\bZ^{i+1}\btheta_t$ for all $t$. If $\bZ^{i+1}$ does not achieve sufficient decrease in $F_2$, the algorithm tests if a proximal gradient step on the original $\bW^i$ would offer a lower total cost than $\bZ^{i+1}$. To this end, the algorithm reconstructs data using $\bW^i$, takes a gradient descent step on $\bW^i$ to find $\bY^i$, and applies the proximal operator on $\bY^i$ to find $\bbZ^{i+1}$ in Steps~\ref{alg:NAPG:step12}--\ref{alg:NAPG:step14}. Step~\ref{alg:NAPG:step13} does not require solving another batch of $T$ OPFs because the OPFs corresponding to $\bW^i$ have already been solved in the previous APG iteration, when the current $\bW^i$ was then termed $\bW^{i+1}$. On the other hand, to evaluate $F_2(\bbZ^{i+1})$ in Step~\ref{alg:NAPG:step15}, the algorithm does have to solve another batch of $T$ OPFs corresponding to data $\bbZ^{i+1}\btheta_t$ for all $t$. Sequence $\alpha_i$ is updated in Step~\ref{alg:NAPG:step17} exactly as in Algorithm~\ref{alg:PGD}. The sequence $c_i$ needed in the sufficient decrease condition of Step~\ref{alg:NAPG:step9} is updated in Steps~\ref{alg:NAPG:step18}--\ref{alg:NAPG:step19}.

Algorithm~\ref{alg:NAPG} solves three batches of $T$ OPFs, one for each one of Steps~\ref{alg:NAPG:step7}, \ref{alg:NAPG:step9}, and \ref{alg:NAPG:step15}. Interestingly, if the OPF is a linear/quadratic program as in~\eqref{eq:opf0}, techniques from multiparametric programming (MPP) can speed up the tasks of finding the minimizers and their sensitivities for a batch of $T$ OPF problem instances. For example, in~\cite{TJKT20}, we were able to speed up the running times of OPF batches by an order of magnitude. Details on sensitivity analysis of the OPF and MPP are omitted as they are quite technical and fall beyond this work's scope.

The next lemma (shown in the appendix) provides a necessary condition on $\lambda_2$ for a critical point of BGL to be zero. 

\begin{lemma}\label{le:lambda2_max}
Let $\bar{\lambda}_2=\max_p \|\bk_p\|$, where $\bk_p$ is the $p$-th column of matrix $\bK=\frac{1}{T}(\nabla_{\btheta} \bx(\bzero))^\top\sum_{t=1}^T (\bx(\bzero)-\bx_t)\btheta_t^\top$ and $\bx(\bzero)$ is the solution of \eqref{eq:opf} for $\btheta=\bzero$. The zero matrix is a critical point of \eqref{eq:bglasso} if and only if $\lambda_2\geq\bar{\lambda}_2$.
\end{lemma}

Similar to GL, we can also pursue a two-stage BGL approach. First, use BGL to select $K$ OPF features and determine a selection matrix $\bS_\text{BGL}$ from~\eqref{eq:distil} based on the sparsity pattern of $\bW_\text{BGL}$. Second, design the reconstruction matrix $\bC$ as:
\begin{equation}
\bC_\text{BGL2}=\arg\min_{\bC}~\frac{1}{2T}\sum_{t=1}^T \|\bx_t-\hbx_t(\bC\bS_\text{BGL}^\top)\|^2.
\end{equation}
The two-stage BGL (BGL2) returns the distillation matrix:
\begin{equation}
\bW_\text{BGL2}=\bC_\text{BGL2}\bS_\text{BGL}^\top.    
\end{equation}

\section{Numerical Tests}\label{sec:tests}
The proposed OPF data distillation approaches were numerically evaluated using a single-phase rendition of the IEEE 37-bus benchmark feeder and {a single-phase 1,136-bus feeder obtained from the IEEE 8,500-bus feeder~\cite{TJKT20}.} Optimal DER reactive power injections $\bx_t$ and $\hbx_t$ were obtained by solving~\eqref{eq:opf} using YALMIP and SeDuMi in MATLAB~\cite{YALMIP,sedumi}. All tests were run on a MacBook laptop computer with an M3 Pro chip and 36 GB of RAM. Type-1 methods took less than a few minutes to run. Type-2 methods took from seconds to a few minutes to run for smaller $K$, and a few hours for larger $K$.

\subsection{Numerical Tests on the 37-bus Feeder}\label{subsec:37}
\emph{Data generation.} Active power demand and solar generation data were collected from the Pecan Street dataset~\cite{pecandata}. We collected minute-based kW load and solar generation data from 25 households between 8 AM and 5 PM during the entire year of 2019. We also collected synthetically generated electric vehicle (EV) profiles over a year from~\cite{Muratori2017}. Among the 37 feeder buses, only 25 correspond to non-zero injections and were simulated to host loads. These 25 buses correspond to medium-voltage nodes, serving multiple residential customers each. To simulate aggregate load profiles, for each one of the medium-voltage nodes, we randomly sampled 5 out of the 25 Pecan Street households and added their profiles. The obtained load profiles were scaled to match twice the kW benchmark load. Because the original dataset does not include reactive power demands, lagging power factors were simulated by randomly drawing them uniformly within $[0.85,1.00]$ across time and nodes. We simulated 10 PVs installed on buses $\{2,4,7,9,11,14,17,20,22,25\}$. Solar profiles from 5 randomly selected Pecan Street households were aggregated to generate the solar profile for each of the 10 medium-voltage nodes. Solar profiles were scaled to match twice the peak kW benchmark load. We randomly sampled 25 EV profiles from~\cite{Muratori2017} and assigned each to the aggregate load profiles. 


\begin{figure}[t]
\centering
\includegraphics[width=0.46\textwidth]{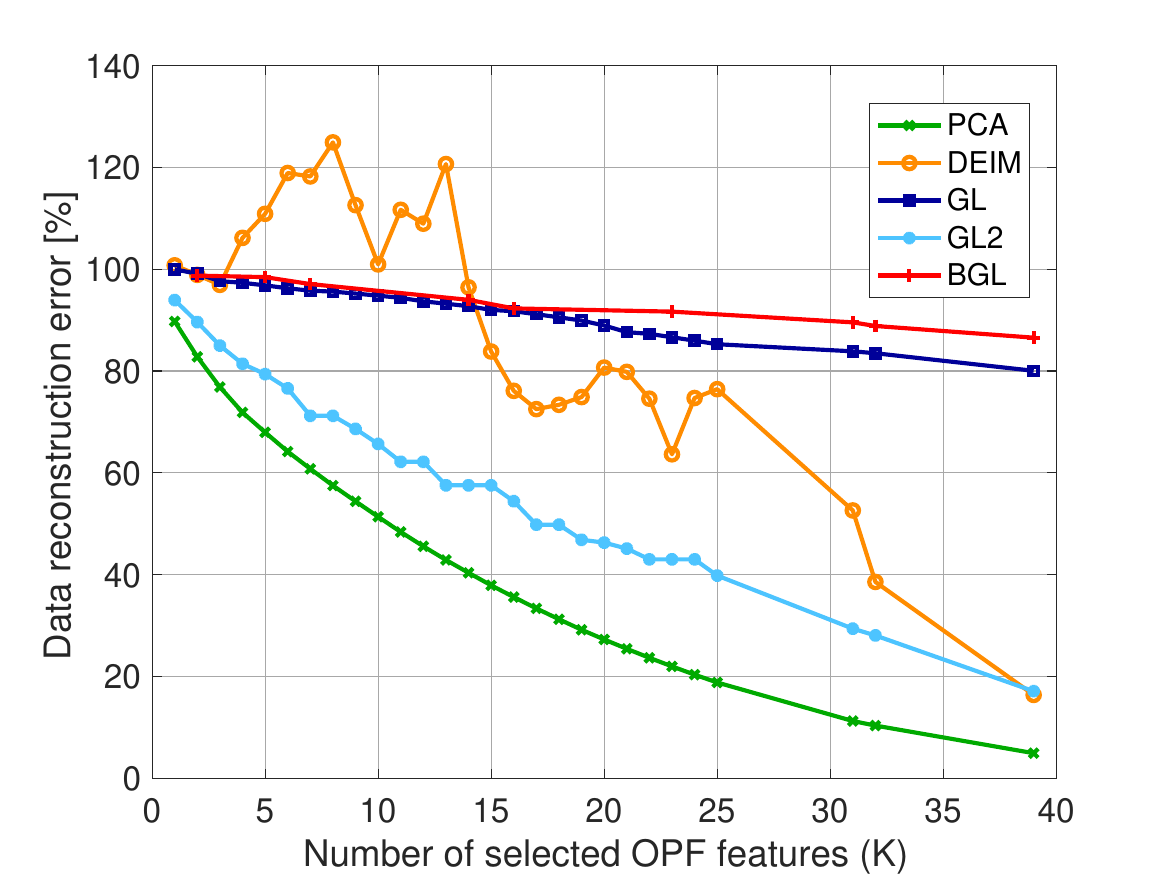} 
\caption{Normalized data reconstruction error $f_1(\bW)/\|\bTheta\|_F^2=\|\bTheta-\hbTheta\|_F^2/\|\bTheta\|_F^2$ obtained by OPF data distillation methods for increasing numbers of OPF features for the single-phase 37-bus system.}
\label{fig:dataerrors}
\end{figure}

\begin{figure}[t]
\centering
\includegraphics[width=0.46\textwidth]{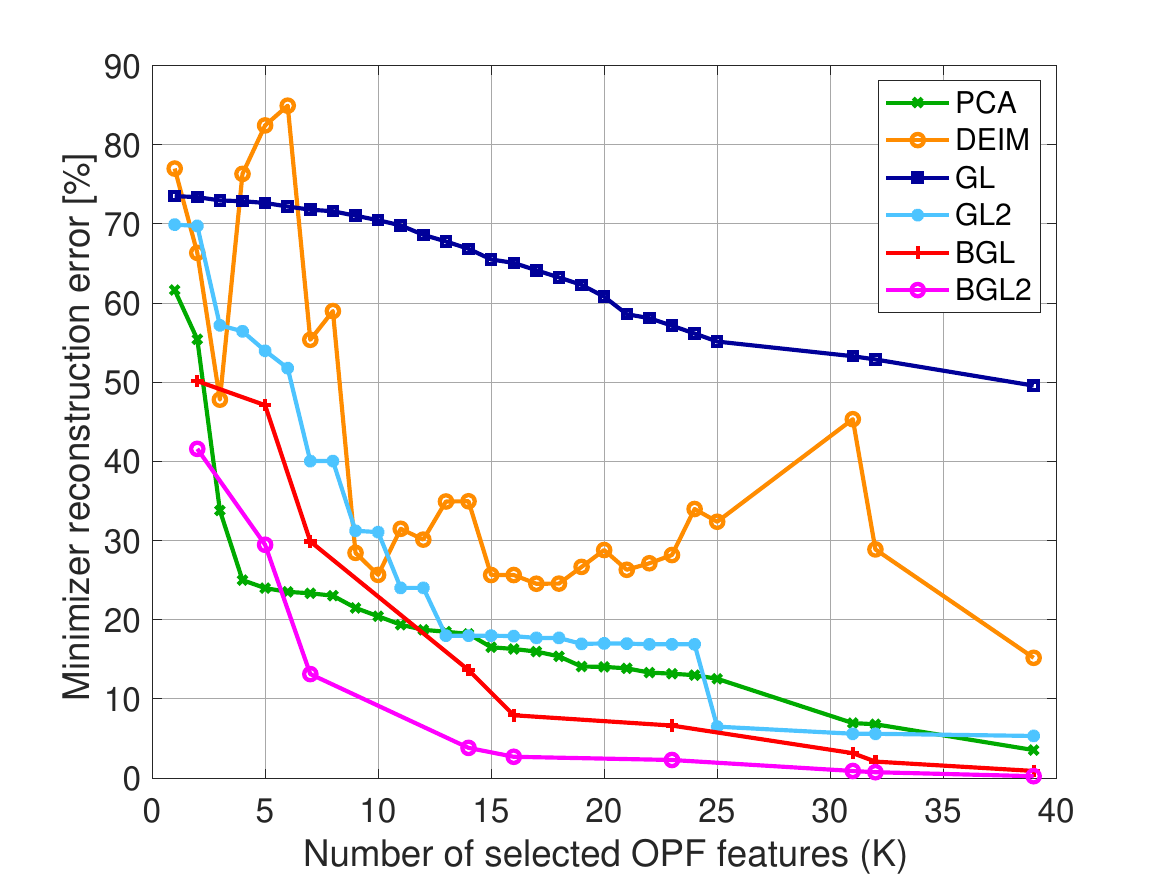} 
\caption{Normalized minimizer error $f_2(\bW)/\bX\|_F^2=\|\bX-\hbX\|_F^2/\|\bX\|_F^2$ obtained by OPF data distillation methods for increasing numbers of OPF features for the single-phase 37-bus system.}
\label{fig:minimizererrors}
\end{figure}

\emph{Fidelity in OPF data under single-phase system.} The first test compares the proposed methods in terms of fidelity in reconstructing OPF data. The test involved four Type-1 approaches (PCA, DEIM, GL, and GL2) and {two} Type-2 approaches (BGL). We used the normalized squared error $\tfrac{f_1(\bW)}{\|\bTheta\|_F^2}$ as a data fidelity metric. This metric was evaluated for an increasing number $K$ of selected OPF features. Data reconstruction errors are expected to decrease for increasing $K$. Parameter $K$ is entered explicitly for PCA and DEIM. For GL-type approaches, we can obtain $K$ non-zero groups in $\bW$ by varying parameters $\lambda_1$ and $\lambda_2$. Figure~\ref{fig:dataerrors} illustrates the data reconstruction error for varying $K$. PCA serves only as a lower bound, as it does not offer data selection. Although DEIM does select data, its performance is unpredictable and exhibits no clear trend. The errors attained by plain GL decrease only marginally as $K$ increases. On the contrary, the GL2 reduces errors more dramatically and follows the trend set by PCA while featuring data selection. BGL behaves similarly to GL and bears no meaningful improvement. {Not shown in Figure~\ref{fig:dataerrors} is the performance curve of BGL2. BGL2 had the worst data reconstruction errors among all methods for most values of $K$, and the errors did not follow any specific trend.} Overall, the GL2 approach is the most effective method for approximating $\bTheta$ when considering sensor selection.

\emph{Fidelity in OPF solutions.} The second test compares four Type-1 approaches to two Type-2 approaches (BGL and BGL2) in terms of fidelity in finding OPF minimizers $\bX$. Figure~\ref{fig:minimizererrors} illustrates the normalized squared error $\tfrac{f_2(\bW)}{\|\bX\|_F^2}$ for increasing $K$. PCA achieves the smallest error across all Type-1 designs, except in a few cases where GL2 yields smaller errors. DEIM does not exhibit a clear trend, and GL shows a slight improvement with increasing $K$. Although the errors attained by GL2, BGL, and BGL2 decrease when $K$ increases, BGL2 achieves smaller errors. The tests corroborate that BGL2 is the preferred method for approximating OPF solutions. {From Figure~\ref{fig:minimizererrors}, it can be seen that when $K\geq 25$, the minimizer errors of GL2 are very close to those obtained by BGL and BGL2. At $K=31$, GL2 achieves a minimizer error of about $6\%$, whereas BGL and BGL2 achieve minimizer errors of about $3\%$ and $1\%$, respectively. Therefore, at larger values of $K$, GL2 may be preferable, as it can achieve small minimization errors while forgoing the increased computational burden of BGL or BGL2. This behavior is expected as discussed in Remark~\ref{re:comparison}.}

\begin{figure}[t]
    \centering
\includegraphics[scale=0.41]{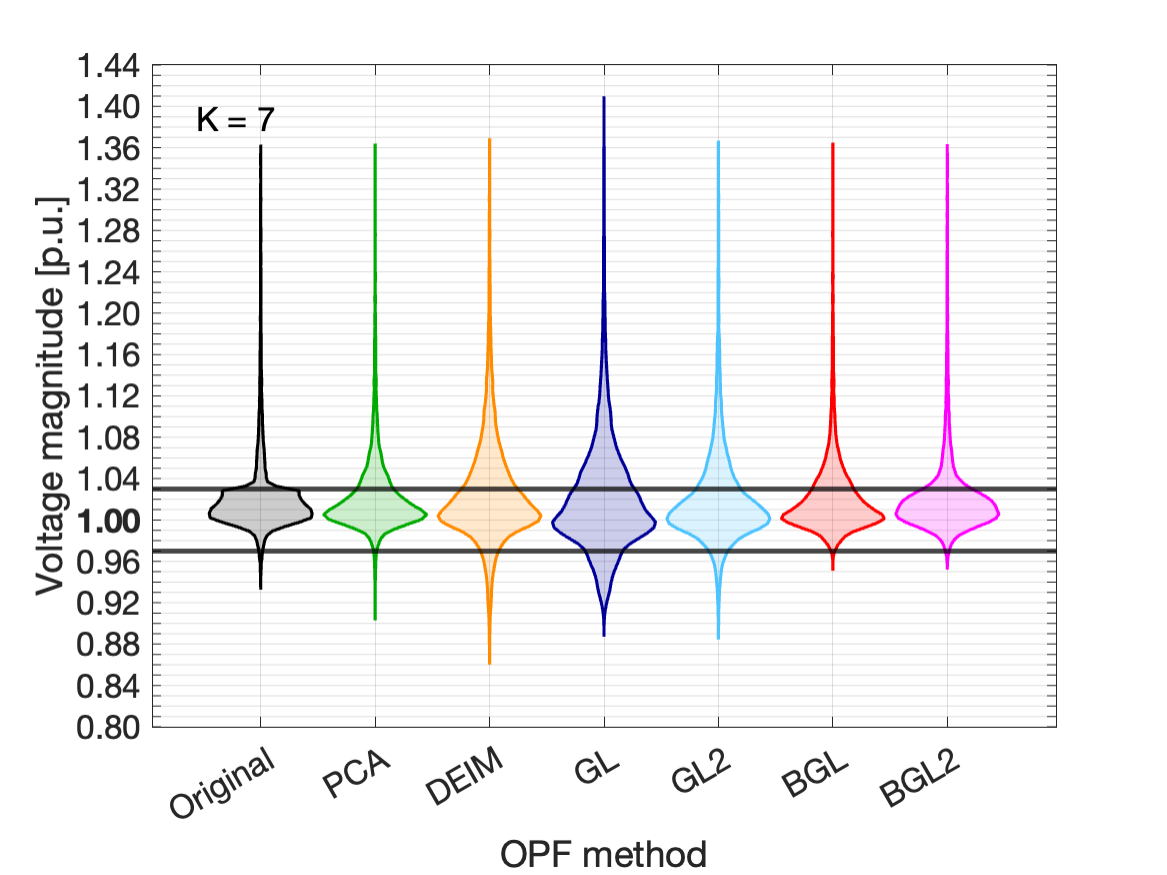}
\includegraphics[scale=0.41]{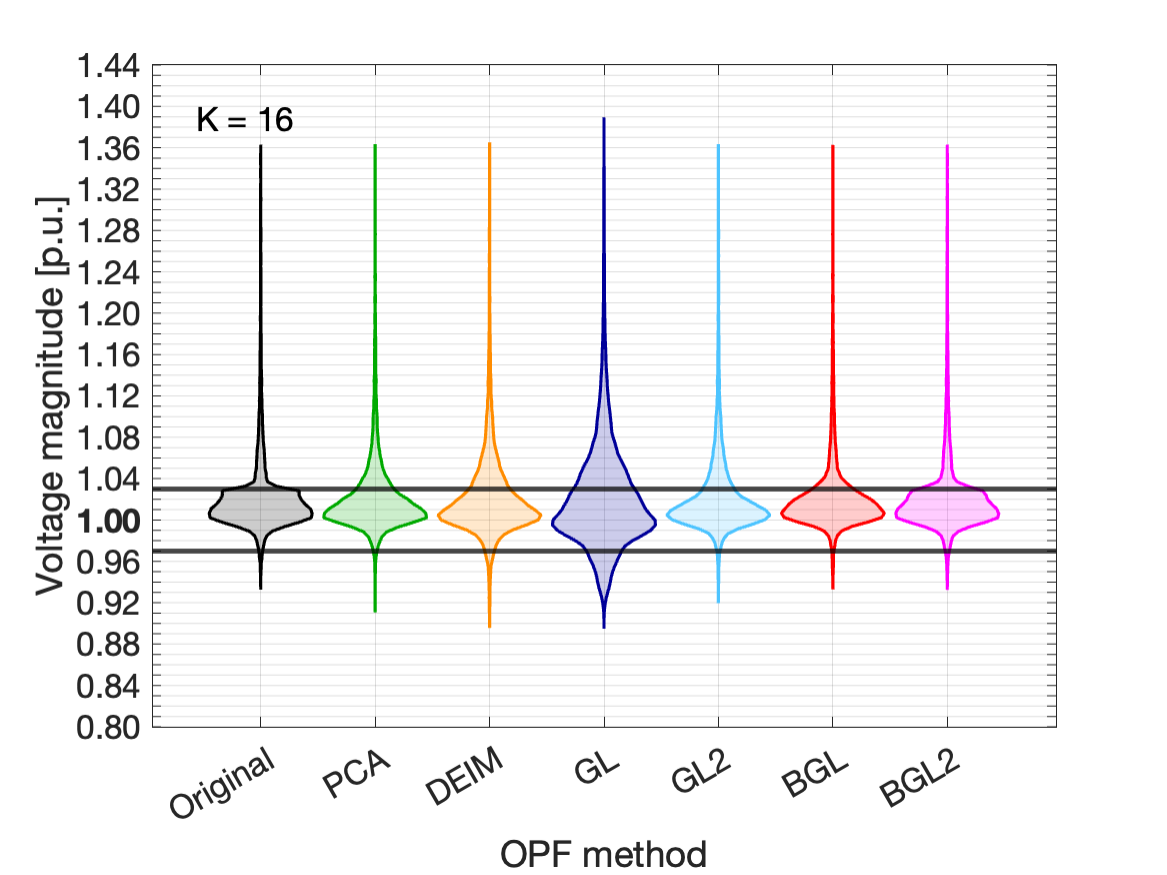}
    \caption{Violin plots of voltage magnitudes computed {using the linearized grid model of \eqref{eq:voltage}} for the IEEE 37-bus single-phase network for $K=7$ (top panel) and $K=16$ (bottom panel) OPF features. Voltages are computed across all buses and scenarios using the \emph{actual} OPF data. {The two black solid horizontal lines mark the $\pm 3\%$ deviations in voltages.} The violin plots obtained using both BGL and two-stage BGL for $K=16$ are almost identical to those of the original OPF method.}
    \label{fig:voltageplots}
\end{figure}

\begin{figure}[t]
    \centering
\includegraphics[scale=0.41]{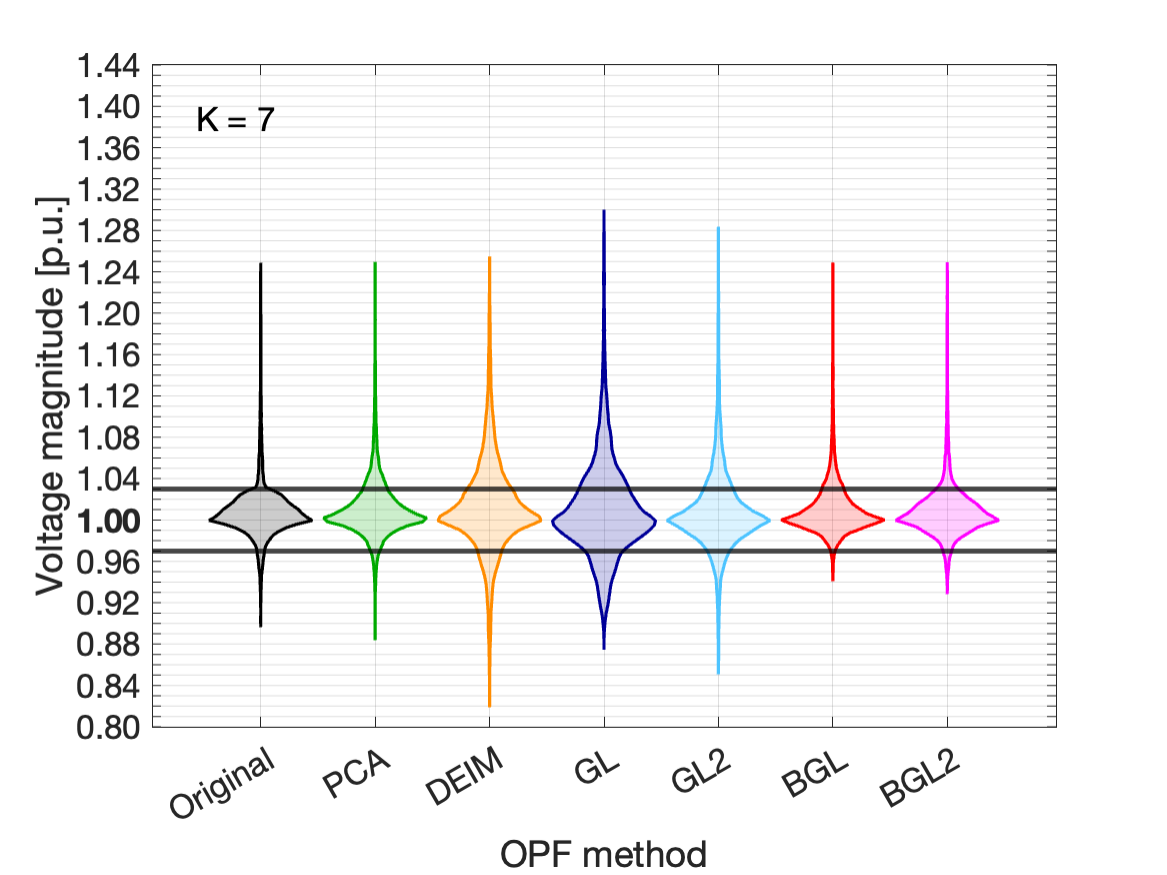}
\includegraphics[scale=0.41]{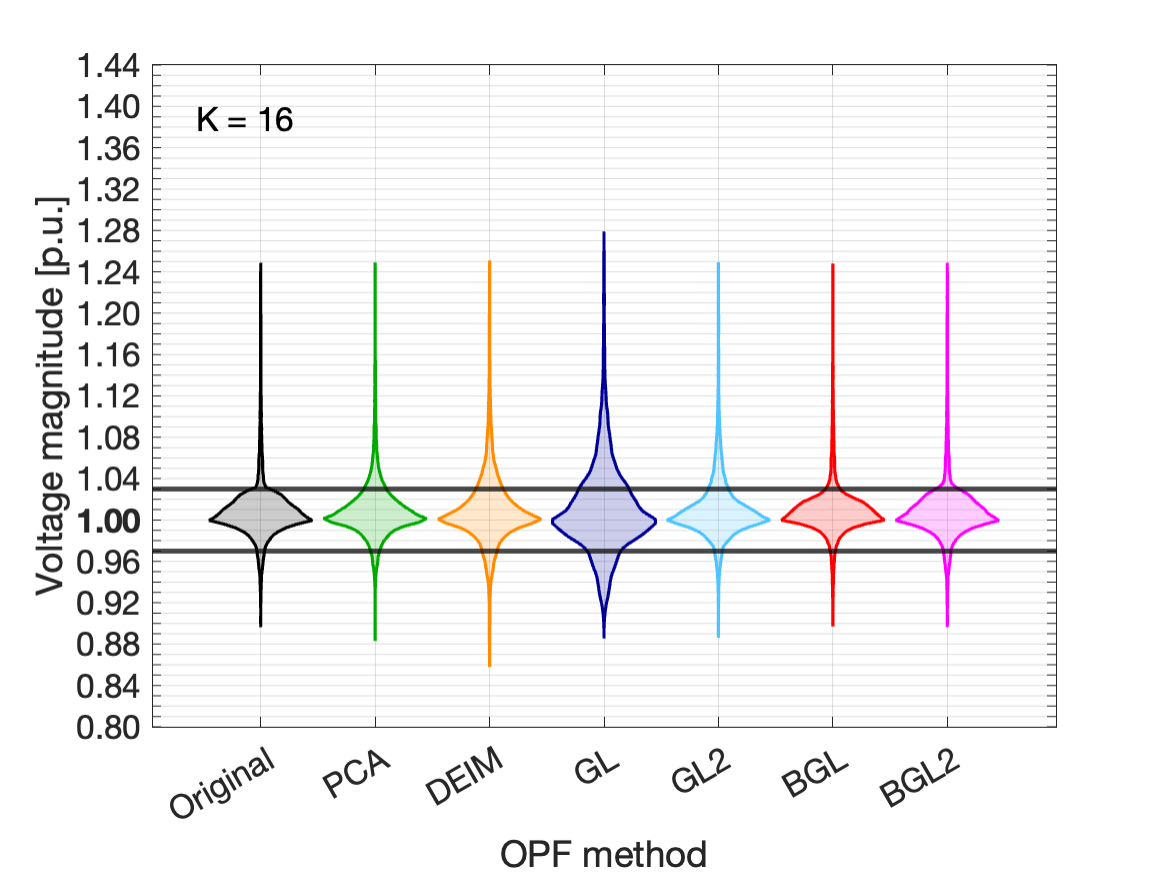}
    \caption{{Violin plots of voltage magnitudes computed using the exact AC grid model for the IEEE 37-bus single-phase network for $K=7$ (top panel) and $K=16$ (bottom panel) OPF features. Voltages are computed across all buses and scenarios using the \emph{actual} OPF data. The two black solid horizontal lines mark the $\pm 3\%$ deviations in voltages. The violin plots obtained using both BGL and two-stage BGL for $K=16$ are almost identical to those of the original OPF method.}}
    \label{fig:acvoltageplots}
\end{figure}

\emph{Feasibility of OPF solutions.} In addition to minimizing $f_2(\bW)$, it is critical that $\hbx_t=\bx(\hbtheta_t)$ does not violate voltage constraints when applied to the grid under actual loading $\btheta_t$. We computed the nodal voltages obtained by all OPF data distillation methods according to {the linearized grid model of~\eqref{eq:voltage}.} Regardless of how DER setpoints were obtained, their voltage effect on the grid was computed using the actual loading conditions. Figure~\ref{fig:voltageplots} depicts the violin plots of all bus voltages across all instances for each method and two values of $K=7$ and $K=16$. Note that even the OPF fed with the original data experiences voltages outside the $\pm3\%$ range because voltage constraints are treated as soft constraints. The tests show that voltage magnitudes are more concentrated under both BGL and BGL2 for both values of $K$, {with a few extreme outliers}. Interestingly, the voltage distributions attained by BGL2 using only $K=16$ OPF features are almost identical to those achieved by the original OPF fed by all $P=50$ features. 

{Figure~\ref{fig:acvoltageplots} depicts violin plots of voltages based on the exact AC grid model. The obtained voltage distributions are similar to those of Fig.~\ref{fig:voltageplots}, and again, BGL2 with $K=16$ provides voltage distributions almost identical to those achieved by the original OPF fed by all $P=50$ OPF features. If the focus is on voltage feasibility rather than fidelity in OPF decisions, BGL/BGL2 may exhibit no significant advantage for larger values of $K$.}

\begin{figure}[t]
\centering
\includegraphics[scale=0.39]{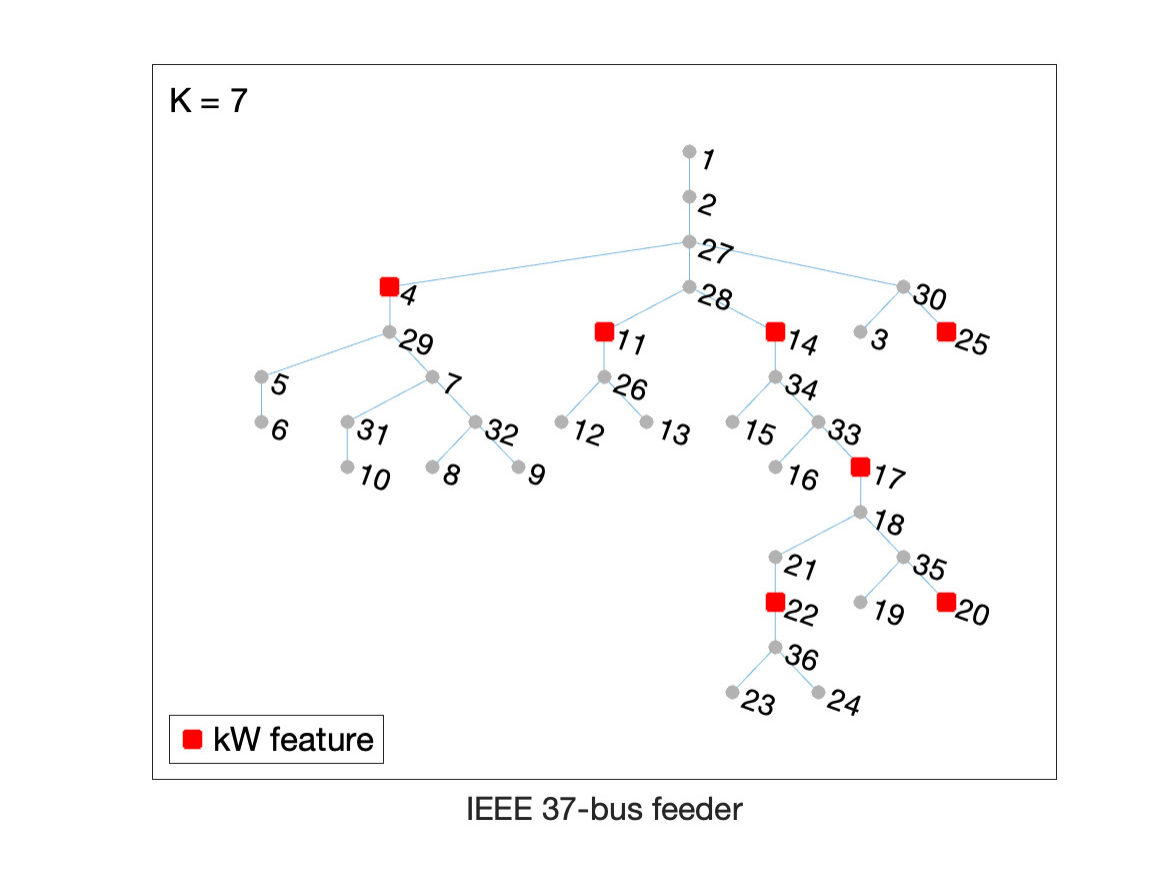}
\includegraphics[scale=0.39]{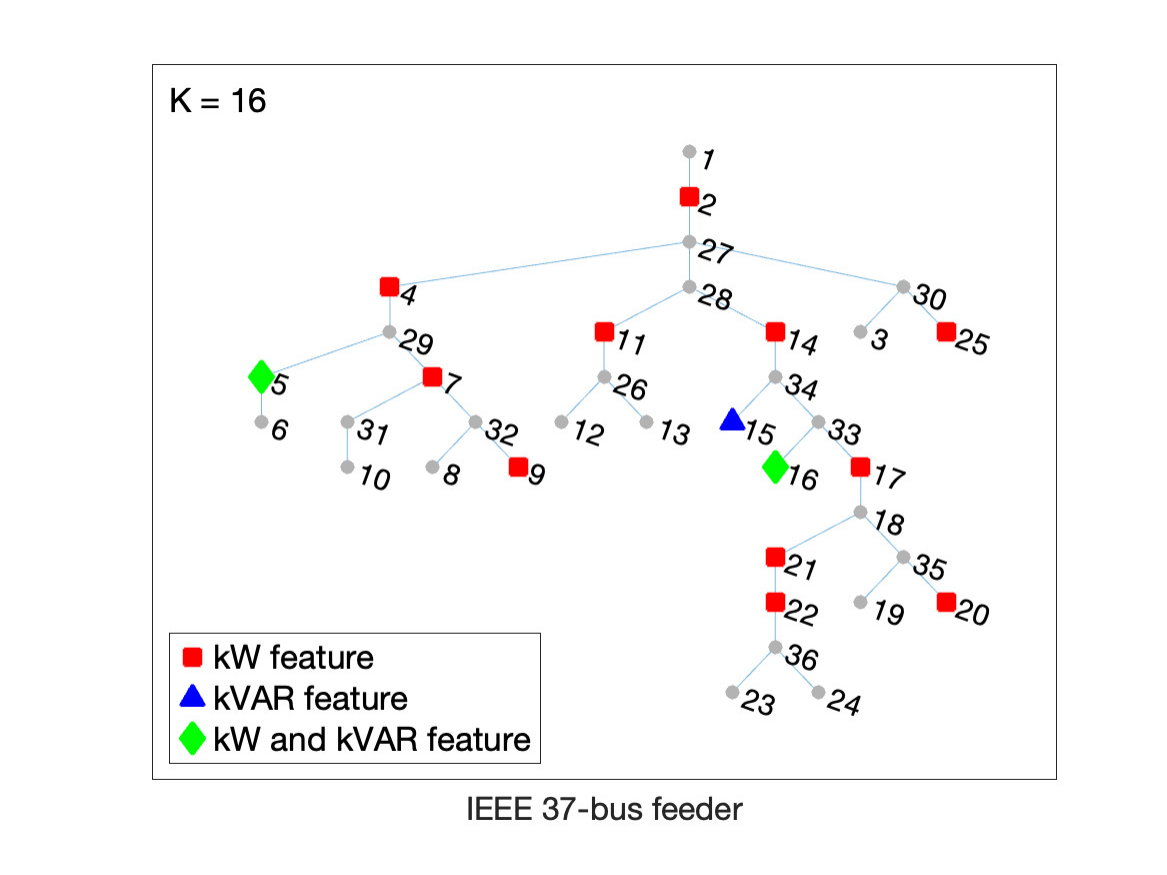}
\vspace*{-1em}
\caption{OPF features selected by BGL for $K=7$ and $K=16$.}
\label{fig:featurelocation}
\end{figure}

\emph{Selected OPF features.} We next observed which features were selected for different values of $K$. Figure~\ref{fig:featurelocation} shows that when $K=7$, only active power demands at seven different nodes were selected. However, when $K=16$, both active and reactive power demands were selected for some nodes, and for other nodes either active or reactive demands were selected.

\emph{Sample size effect.} We next evaluated how using different sample sizes affected our proposed methods. Fifty (50) scenarios were randomly sampled from the $T=800$ scenarios that were used in previous tests. The same $\lambda_2$ values resulted in different values of $K$. For the same number of selected features $K=5$, different features were selected for the two scenarios. The minimizer errors for all Type-1 and Type-2 methods are lower when the number of OPF scenarios is smaller. Data reconstruction errors are also lower when the OPF data has fewer scenarios. The tests show that the number of OPF scenarios affects both data and minimizer reconstruction errors and also decides which OPF features are selected.

\subsection{{Numerical Tests on the 1,136-bus Feeder}}\label{subsec:1136}
\emph{Data generation.} To test the scalability of the proposed methods, we also ran tests on a single-phase 1,136-bus feeder obtained from the IEEE 8,500-bus test feeder by keeping buses sited below bus 338~\cite{TJKT20}. Out of its 1,136 nodes, only 519 host non-zero injections. We obtained active power demand and solar generation data at one-minute intervals from the Smart* project~\cite{Barker2012sustkdd}. The dataset contained active loads from 444 homes and solar generation from 45 PVs between 5:00 AM and 9:30 PM. We aggregated load demands from multiple homes to simulate 519 active load profiles for the 519 non-zero injection nodes of the tested feeder. Load profiles were scaled to match the kW benchmark load. Reactive load profiles were simulated by drawing power factors uniformly at random within $[0.85,0.99]$ across time and nodes. We simulated 20 PVs by aggregating the solar generation from the 45 PVs from~\cite{Barker2012sustkdd} and randomly installed them on 20 of the non-zero injection nodes (514, 535, 589, 600, 601, 616, 656, 664, 707, 730, 781, 839, 840, 897, 914, 959, 981, 1036, 1058 and 1069).

\begin{table}[t]
\centering
{\caption{Fidelity in OPF solutions for the IEEE 1,136-bus system}
\label{tbl:OPF_error_1136_bus}}
\vspace*{-.5em}

\begin{tabular}{|c|c|c|c|c|c|}
\hline
\multirow{2}{*}{$K$ out of $P=1,038$} & \multicolumn{5}{c|}{Normalized error in OPF solutions [\%]} \\ \cline{2-6}
& \textbf{PCA} & \textbf{GL} & \textbf{GL2} & \textbf{BGL} & \textbf{BGL2} \\ \hline\hline
2 & 99.29 & 99.99 & 99.65 & 88.55 & 40.22 \\ \hline
5 & 98.29 & 99.99 & 99.11 & 53.54 & 36.30 \\ \hline
21 & 93.28 & 99.96 & 96.33& 39.95 & 28.35 \\ \hline
36 & 88.86 & 99.92 & 93.80 & 20.69 & 6.02 \\ \hline
116 & 68.98 & 99.77 & 80.91 & 18.11 & 4.91 \\ \hline
\end{tabular}
\vspace*{-1em}
\end{table}

{\emph{Fidelity.} The first test for the 1,136-bus network compares GL/GL2 and BGL/BGL2 in terms of reconstructing the OPF minimizers $\bX$. Table~\ref{tbl:OPF_error_1136_bus} shows the normalized squared errors for 5 values of $K$. Similar to the smaller 37-bus network, PCA attains the smallest errors among all Type-1 designs in the 1,136-bus network. Again, BGL2 is the best method for approximating OPF solutions. The reported errors verify the importance of the two-stage approach.}

\begin{figure}[t]
    \centering
\includegraphics[scale=0.4]{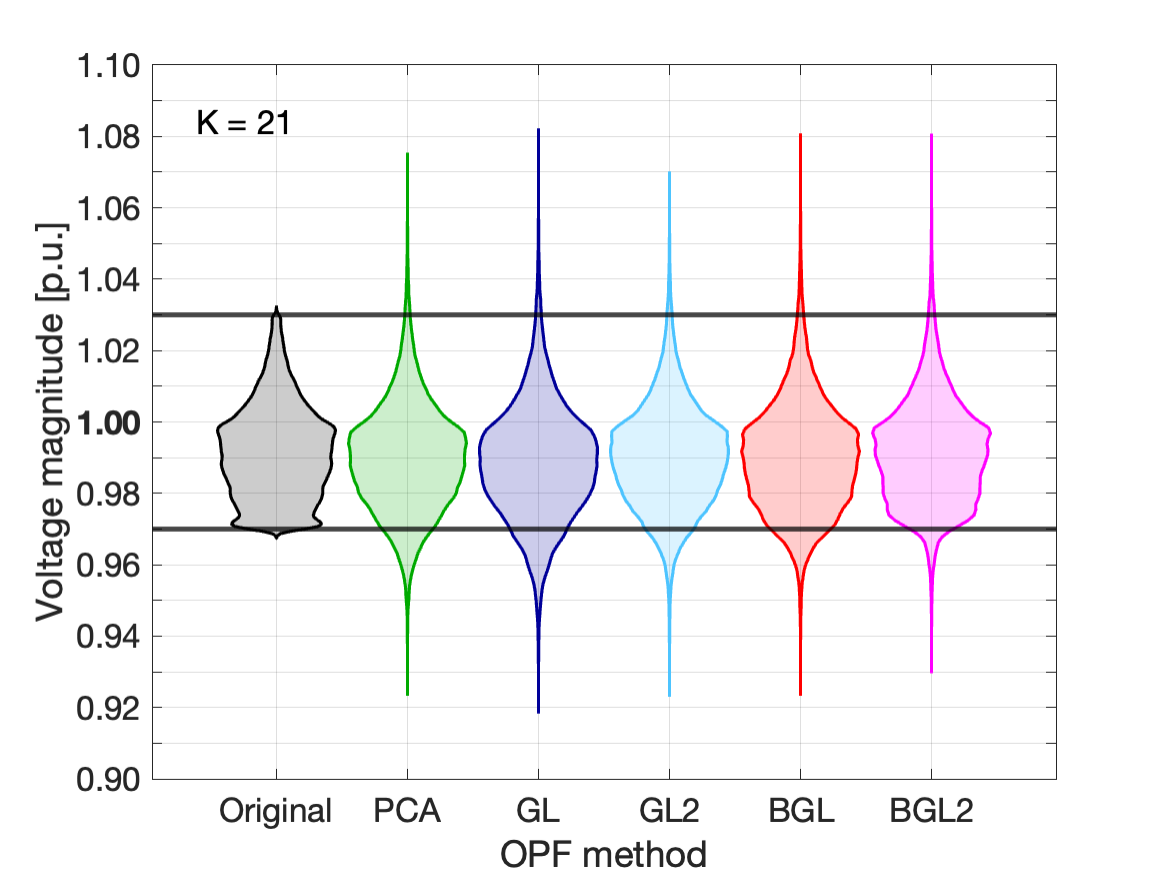}
\includegraphics[scale=0.4]{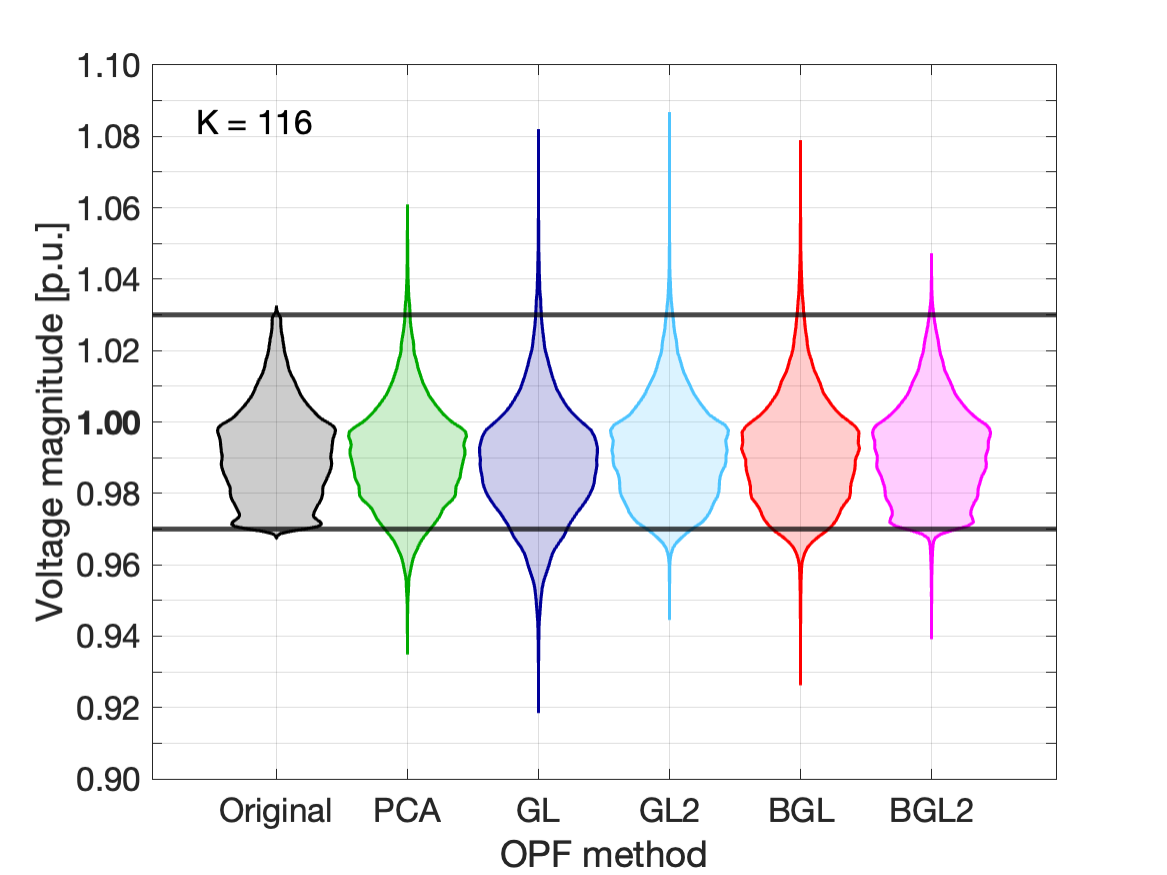}
    \caption{{Violin plots of voltage magnitudes for the single-phase 1,136-bus network experienced under all OPF methods for $K=21$ (top) and $K=116$ (bottom) out of 1,038 OPF features. Voltages are computed across all buses and scenarios using the linearized grid model of \eqref{eq:voltage} and \emph{actual} OPF data. The two black solid horizontal lines mark the $\pm 3\%$ deviations in voltages. The violin plots obtained using BGL2 with $K=116$ are the closest to those obtained from the original OPF method.}}
    \label{fig:voltageplots2}
\end{figure}

\begin{figure}[t]
    \centering
\includegraphics[scale=0.4]{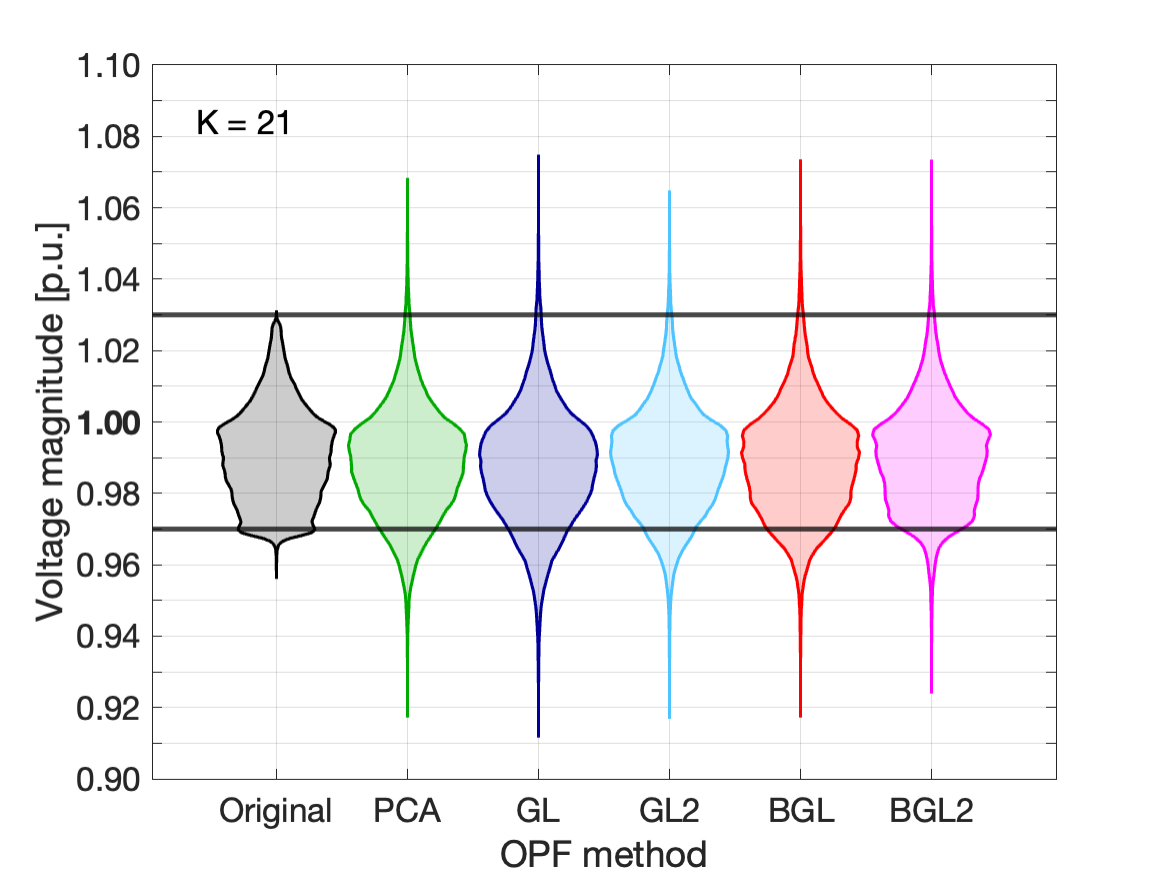}
\includegraphics[scale=0.4]{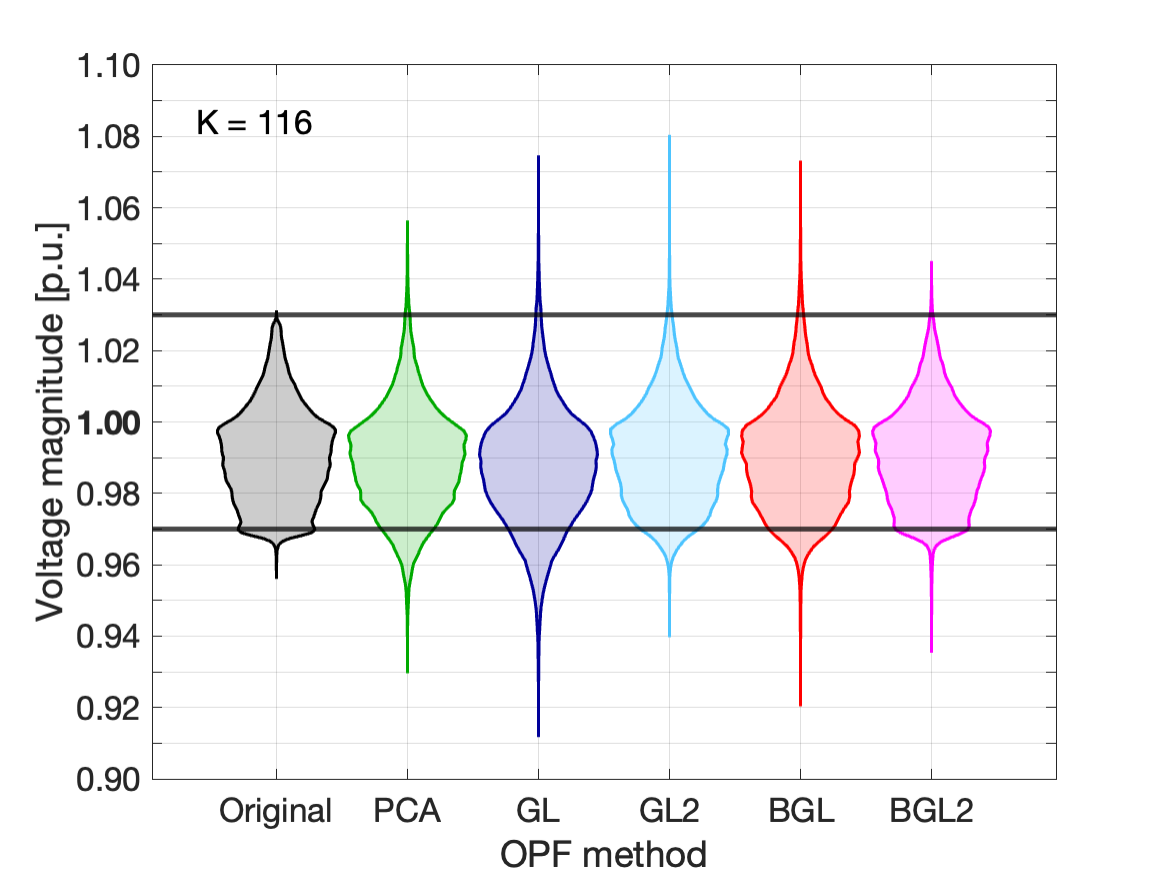}
    \caption{{Violin plots of voltage magnitudes computed using the exact AC grid model for the single-phase 1,136-bus network experienced under all OPF methods for $K=21$ (top) and $K=116$ (bottom) out of 1,038 OPF features. Voltages are computed across all buses and scenarios using the linearized grid model of \eqref{eq:voltage} and \emph{actual} OPF data. The violin plots obtained using BGL2 with $K=116$ are the closest to those obtained from the original OPF method.}}
    \label{fig:acvoltageplots2}
\end{figure}

{\emph{Feasibility.} We simulated the approximate OPF solutions along with the true loading conditions to compute nodal voltages per~\eqref{eq:voltage}. Figure~\ref{fig:voltageplots2} shows the violin plots across all nodes and scenarios for all methods. The tests show that the voltage magnitudes are more concentrated for BGL2 and that by using only $K=116$ OPF features, we can achieve voltage distributions almost identical to those achieved by the original OPF fed by all $P=1,038$ OPF features.} {Figure~\ref{fig:acvoltageplots2} depicts violin plots of voltages based on the exact AC grid model. Conclusions similar to those of Fig.~\ref{fig:voltageplots2} can be drawn.}

\color{black}

\section{Conclusions}\label{sec:conclusions}
We have proposed an OPF data distillation framework to improve the fidelity of either OPF features or OPF minimizers. Type-2 OPF data distillation yields a non-convex program, which is solved by a proximal gradient algorithm. To alleviate the computational burden of finding the minimizers and their sensitivities for a batch of OPF instances, we have leveraged results from multiparametric programming. 

{Extensive numerical tests using real-world data on 37-bus and 1,136-bus single-phase feeders demonstrate the advantages of Type-2 over Type-1 methods in approximating the minimizer and satisfying network constraints at the expense of increased complexity. Numerical tests have shown that optimal DER schedules can be approximated reasonably well upon collecting only $K=16$ rather than $P=50$ OPF features for the IEEE 37-bus single-phase feeder. For the larger system, DER schedules can be approximated within 5\% error using $K=116$ out of $P=1,038$ features while nearly matching the voltage distribution of the exact OPF solutions. For larger values of compression ratios $K/P$, the advantage of Type-2 methods diminishes in general, and Type-1 methods perform reasonably well.}

The proposed framework sets a solid foundation for exciting research directions: \emph{i)} Extending data distillation to AC-OPF formulations or other OPF variants; \emph{ii} Pursuing stochastic variants of APGD that process one or a few OPF instances per iteration to reduce the computational cost; \emph{iii)} Leveraging OPF data distillation tools towards optimal meter placement to support OPF operations; \emph{iv)} Upgrading the linear to a nonlinear reconstruction module (deep neural network); \emph{v)} Pursuing unsupervised counterparts of OPF data distillation to waive the need of generating a labeled OPF dataset; and \emph{vi)} Identifying the most influential OPF features so that their communication to the DSO is protected against cyberattacks.

\section*{Appendix}
\begin{IEEEproof}[Proof of Lemma~\ref{le:lambda_max}] The subdifferential of $g(\bW)$ with respect to the $p$-th column of $\bW$ is
\begin{equation}
\label{eq:subdiff}
\partial_{\bw_p}g(\bW) =
\begin{cases}
\frac{\bw_p}{\|\bw_p\|} &,~\text{if } \bw_p \neq \bzero \\
\left\{\bg_p\in\mathbb{R}^P: \|\bg_p\| \leq 1\right\} &,~\text{if } \bw_p = \bzero
\end{cases}
\end{equation}
for all $p$. Function $g(\bW)$ is differentiable with respect to $\bw_p$ for all $\bw_p\neq \bzero$. Its subdifferential at $\bw_p=\bzero$ can be defined as the set of all vectors with an $\ell_2$-norm not exceeding one. 

Optimality conditions for subdifferentiable functions predicate that a matrix $\bW_\text{GL}$ is a minimizer of \eqref{eq:glasso} if and only if the zero matrix belongs to the subdifferential of $f_1+\lambda_1 g$ evaluated at $\bW_\text{GL}$. Therefore, the minimizer of~\eqref{eq:glasso} satisfies:
\begin{equation*}
\bzero \in \nabla_{\bW}f_1(\bW_\text{GL}) + \lambda_1 \partial_{\bW} g(\bW_\text{GL}).
\end{equation*}
Substituting $\nabla_{\bW}f_1(\bW)$ from \eqref{eq:gradient} above yields
\begin{equation}\label{eq:oc1}
\bzero \in (\bW_\text{GL}-\bI)\bC_\theta+\lambda_1 \partial_{\bW} g(\bW_\text{GL}).
\end{equation}

If $\bW_\text{GL}=\bzero$, the optimality condition of \eqref{eq:oc1} provides:
\begin{equation*}
\lambda_1 \begin{bmatrix} \bg_1 & \bg_2 & \cdots & \bg_P \end{bmatrix} = \bC_{\theta}\quad\quad  \text{or}\quad\quad  \lambda_1 \bg_p = \bc_p~~\forall p
\end{equation*}
where $\bg_p$ is defined by the lower branch of \eqref{eq:subdiff} as it evaluates $\partial_{\bw_p}g$ at $\bW_\text{GL}=\bzero$. Since $\|\bg_p\|\leq 1$ for all $p$ by the definition of the subgradient at zero, it follows that $\|\bc_p\|=\lambda_1 \|\bg_p\|\leq \lambda_1$ for all $p$. This establishes that $\lambda_1\geq \bar{\lambda}_1$. Conversely, if $\lambda_1\geq \bar{\lambda}_1$, matrix $\bW_\text{GL}=\bzero$ can be shown to satisfy \eqref{eq:oc1} by selecting $\bg_p=\frac{1}{\lambda_1}\bc_p$ so that $\|\bg_p\|_2\leq 1$ for all $p$. 
\end{IEEEproof}




\begin{IEEEproof}[Proof of Lemma~\ref{le:gradient}]
We first compute the gradient of $f_2$ defined in \eqref{eq:f2} with respect to vector $\bw=\vectorize(\bW)$. This is a $P^2$-long vector wherein all columns of $\bW$ are stacked vertically. We will use the property that for any three matrices $(\bA,\bB,\bC)$ of conformable dimensions, it holds that:
\begin{equation}\label{eq:vecproperty}
\vectorize(\bA\bB\bC)=(\bC^\top\otimes \bA)\vectorize(\bB)
\end{equation}
where $\otimes$ is the Kronecker product.

The gradient of the $t$-th summand of $f_2$ is
\[\nabla_{\bw}\|\bx_t-\hbx_t\|^2=2(\nabla_{\bw}\hbx_t)^\top(\hbx_t-\bx_t).\]
The Jacobian $\nabla_{\bw}\hbx_t$ can be found using the chain rule as
\[\nabla_{\bw}\hbx_t=\nabla_{\hbtheta_t}\hbx_t\cdot \nabla_{\bw}\hbtheta_t.\]
Thanks to~\eqref{eq:vecproperty}, we can express data $\hbtheta_t=\bW\btheta_t$ as
\[\hbtheta_t=\vectorize(\hbtheta_t)=\vectorize(\bW\btheta_t)=(\btheta_t^\top\otimes \bI_P)\bw\]
from which it follows that $\nabla_{\bw}\hbtheta_t=\nabla_{\bw}\hbtheta_t=\btheta_t^\top\otimes \bI_P$. Substituting the latter into the gradient of $\|\bx_t-\hbx_t\|^2$ yields:
\[\nabla_{\bw}\|\bx_t-\hbx_t\|^2=2(\btheta_t\otimes \bI_P)(\nabla_{\hbtheta_t}\hbx_t)^\top(\hbx_t-\bx_t).\]
Therefore, the gradient of $\|\bx_t-\hbx_t\|^2$ with respect to $\bW$ is
\[\nabla_{\bW}\|\bx_t-\hbx_t\|^2=2(\nabla_{\hbtheta_t}\hbx_t)^\top(\hbx_t-\bx_t)\btheta_t^\top.\]
This can be verified using~\eqref{eq:vecproperty} for $\bA=2(\nabla_{\hbtheta_t}\hbx_t)^\top(\hbx_t-\bx_t)$, $\bB=1$, and $\bC=\btheta^\top$, which shows that
\[\vectorize(\nabla_{\bW}\|\bx_t-\hbx_t\|^2)=\nabla_{\bw}\|\bx_t-\hbx_t\|^2.\]

Summing up the gradients of all $\|\bx_t-\hbx_t\|^2$ across $t$ and dividing by $2T$ completes the proof of this lemma.
\end{IEEEproof}

\begin{IEEEproof}[Proof of Lemma~\ref{le:lambda2_max}] If $\bW_\text{BGL}$ is a critical point, then 
\begin{equation}\label{eq:Type2-gradient}
\bzero \in \nabla_{\bW}f_2(\bzero_{P\times P}) + \lambda_2\partial g(\bzero_{P\times P}).
\end{equation}
The gradient of $f_2$ can be found from Lemma~\ref{le:gradient}. For $\bW_\text{BGL}=\bzero$, it holds $\hbtheta_t=\bW_\text{BGL}\btheta_t=\bzero$ for all $t$. We therefore have that $\hbx_t=\bx(\bzero)$ and $\nabla_{\hbtheta_t}\hbx_t=\nabla_{\btheta}\bx(\bzero)$ for all $t$. The rest of the proof follows the proof of Lemma~\ref{le:lambda_max} and is omitted.
\end{IEEEproof}

\balance
\bibliographystyle{IEEEtran}
\bibliography{myabbrv,power,kekatos}
\end{document}

%% file: stylesV2.tex
\DeclareMathOperator{\rank}{rank}

\DeclareMathOperator{\vectorize}{vec}

\newtheorem{lemma}{Lemma}

\newtheorem{remark}{Remark}

\newcommand \bzero{\mathbf{0}}
\newcommand \bone{\mathbf{1}}

\newcommand \bc{\mathbf{c}}

\newcommand \bg{\mathbf{g}}

\newcommand \bk{\mathbf{k}}

\newcommand \bp{\mathbf{p}}
\newcommand \bq{\mathbf{q}}

\newcommand \bv{\mathbf{v}}
\newcommand \bw{\mathbf{w}}
\newcommand \bx{\mathbf{x}}
\newcommand \by{\mathbf{y}}
\newcommand \bz{\mathbf{z}}
\newcommand \bA{\mathbf{A}}
\newcommand \bB{\mathbf{B}}
\newcommand \bC{\mathbf{C}}

\newcommand \bI{\mathbf{I}}

\newcommand \bK{\mathbf{K}}

\newcommand \bR{\mathbf{R}}
\newcommand \bS{\mathbf{S}}

\newcommand \bU{\mathbf{U}}

\newcommand \bW{\mathbf{W}}
\newcommand \bX{\mathbf{X}}
\newcommand \bY{\mathbf{Y}}
\newcommand \bZ{\mathbf{Z}}

\newcommand \btheta{\boldsymbol{\theta}}

\newcommand \bphi{\boldsymbol{\phi}}

\newcommand \bTheta{\mathbf{\Theta}}

\newcommand \bLambda{\mathbf{\Lambda}}

\newcommand \mcD{\mathcal{D}}

\newcommand \mcO{\mathcal{O}}

\newcommand \tbtheta{\tilde{\boldsymbol{\theta}}}

\newcommand \hbx{\hat{\mathbf{x}}}

\newcommand \hbX{\hat{\mathbf{X}}}

\newcommand \hbtheta{\hat{\boldsymbol{\theta}}}

\newcommand \hbTheta{\hat{\mathbf{\Theta}}}

\newcommand \bbq{\bar{\mathbf{q}}}

\newcommand \bbv{\bar{\mathbf{v}}}

\newcommand \bbx{\bar{\mathbf{x}}}
\newcommand \bby{\bar{\mathbf{y}}}
\newcommand \bbz{\bar{\mathbf{z}}}

\newcommand \bbW{\bar{\mathbf{W}}}

\newcommand \bbY{\bar{\mathbf{Y}}}
\newcommand \bbZ{\bar{\mathbf{Z}}}

\newcommand \bbtheta{\bar{\boldsymbol{\theta}}}

\newcommand \bbTheta{\bar{\mathbf{\Theta}}}

